\documentclass[dvips]{acta}
\usepackage{supertabular,lscape,epsfig}
\usepackage{amssymb}
\usepackage{amsmath}
\usepackage{multirow}
\mathchardef\mhyphen="2D
\usepackage{placeins}
\DeclareSymbolFont{ppa}{OT1}{ppl}{m}{it}
\DeclareMathSymbol{\vv}{\mathalpha}{ppa}{'166}

\SetPages{1}{35}

\SetVol{67}{2017}

\begin{document}
\newcommand\pvalue{\mathop{p\mhyphen {\rm value}}}
\newcommand{\TabApp}[2]{\begin{center}\parbox[t]{#1}{\centerline{
  {\bf Appendix}}
  \vskip2mm
  \centerline{\small {\spaceskip 2pt plus 1pt minus 1pt T a b l e}
  \refstepcounter{table}\thetable}
  \vskip2mm
  \centerline{\footnotesize #2}}
  \vskip3mm
\end{center}}

\newcommand{\TabCapp}[2]{\begin{center}\parbox[t]{#1}{\centerline{
  \small {\spaceskip 2pt plus 1pt minus 1pt T a b l e}
  \refstepcounter{table}\thetable}
  \vskip2mm
  \centerline{\footnotesize #2}}
  \vskip3mm
\end{center}}

\newcommand{\TTabCap}[3]{\begin{center}\parbox[t]{#1}{\centerline{
  \small {\spaceskip 2pt plus 1pt minus 1pt T a b l e}
  \refstepcounter{table}\thetable}
  \vskip2mm
  \centerline{\footnotesize #2}
  \centerline{\footnotesize #3}}
  \vskip1mm
\end{center}}

\newcommand{\MakeTableApp}[4]{\begin{table}[p]\TabApp{#2}{#3}
  \begin{center} \TableFont \begin{tabular}{#1} #4 
  \end{tabular}\end{center}\end{table}}

\newcommand{\MakeTableSepp}[4]{\begin{table}[p]\TabCapp{#2}{#3}
  \begin{center} \TableFont \begin{tabular}{#1} #4 
  \end{tabular}\end{center}\end{table}}

\newcommand{\MakeTableee}[4]{\begin{table}[htb]\TabCapp{#2}{#3}
  \begin{center} \TableFont \begin{tabular}{#1} #4
  \end{tabular}\end{center}\end{table}}

\newcommand{\MakeTablee}[5]{\begin{table}[htb]\TTabCap{#2}{#3}{#4}
  \begin{center} \TableFont \begin{tabular}{#1} #5 
  \end{tabular}\end{center}\end{table}}

\newfont{\bb}{ptmbi8t at 12pt}
\newfont{\bbb}{cmbxti10}
\newfont{\bbbb}{cmbxti10 at 9pt}
\newcommand{\uprule}{\rule{0pt}{2.5ex}}
\newcommand{\douprule}{\rule[-2ex]{0pt}{4.5ex}}
\newcommand{\dorule}{\rule[-2ex]{0pt}{2ex}}
\begin{Titlepage}
\Title{OGLE-ing the Magellanic System: Three-Dimensional Structure\\ of the Clouds and the Bridge using RR~Lyrae Stars}
\Author{A.\,M.~~J~a~c~y~s~z~y~n~-~D~o~b~r~z~e~n~i~e~c~k~a$^1$,~~ 
D.\,M.~~S~k~o~w~r~o~n$^1$,~~ 
P.~~M~r~ó~z$^1$,\\
I.~~S~o~s~z~y~ñ~s~k~i$^1$,~~
A.~~U~d~a~l~s~k~i$^1$,~~ 
P.~~P~i~e~t~r~u~k~o~w~i~c~z$^1$,~~ 
J.~~S~k~o~w~r~o~n$^1$,\\
R.~~P~o~l~e~s~k~i$^{1,2}$,~~
S.~~~K~o~z~³~o~w~s~k~i$^1$,~~ 
£.~~W~y~r~z~y~k~o~w~s~k~i$^1$,~~ 
M.~~P~a~w~l~a~k$^1$,\\
M.\,K.~~S~z~y~m~a~ñ~s~k~i$^1$~~ 
and~~ K.~~U~l~a~c~z~y~k$^3$}
{$^1$Warsaw University Observatory, Al. Ujazdowskie 4, 00-478 Warszawa, Poland\\
e-mail: ajacyszyn@astrouw.edu.pl \\
$^2$Department of Astronomy, Ohio State University, 140 W. 18th Ave., Columbus,\\ OH 43210, USA\\
$^3$Department of Physics, University of Warwick, Gibbet Hill Road,\\ Coventry, CV4 7AL, UK
}
\Received{November 29, 2016}
\end{Titlepage}

\Abstract{We present a three-dimensional analysis of a sample of 
22\,859 type ab RR~Lyr stars in the Magellanic System from the OGLE-IV
Collection of RR~Lyr stars. The distance to each object was
calculated based on its photometric metallicity and a theoretical
relation between color, absolute magnitude and metallicity.

The LMC RR~Lyr distribution is very regular and does not show any
substructures. We demonstrate that the bar found in previous studies
may be an overdensity caused by blending and crowding effects. The
halo is asymmetrical with a higher stellar density in its
north-eastern area, which is also located closer to us. Triaxial
ellipsoids were fitted to surfaces of a constant number
density. Ellipsoids farther from the LMC center are less elongated and
slightly rotated toward the SMC. The inclination and position angle
change significantly with the $a$ axis size. The median axis ratio is
$1:1.23:1.45$.

The RR~Lyr distribution in the SMC has a very regular, ellipsoidal
shape and does not show any substructures or asymmetries. All triaxial
ellipsoids fitted to surfaces of a constant number density have
virtually the same shape (axis ratio) and are elongated along the
line-of-sight. The median axis ratio is $1:1.10:2.13$. The inclination
angle is very small and thus the position angle is not well defined.

We present the distribution of RR~Lyr stars in the Magellanic Bridge
area, showing that the Magellanic Clouds' halos overlap.

A comparison of the distributions of RR~Lyr stars and Classical
Cepheids shows that the former are significantly more spread and
distributed regularly, while the latter are very clumped and form
several distinct substructures.} {Stars: fundamental parameters --
Stars: variables: RR~Lyrae -- Magellanic Clouds -- Galaxies: statistics
-- Galaxies: structure}

\Section{Introduction}
The Magellanic System consists of the Large Magellanic Cloud (LMC) and
Small Magellanic Cloud (SMC) along with a few structures that were
formed as a result of the Clouds' interactions. These structures are:
the Magellanic Stream, the Leading Arm, and the Magellanic Bridge
(MBR) (Gardiner \etal 1994, Gardiner and Noguchi 1996, Yoshizawa and
Noguchi 2003, Connors \etal 2006, R\r{u}\v{z}i\v{c}ka \etal 2009,
2010, Besla \etal 2010, 2012, Diaz and Bekki 2011, 2012,
Guglielmo \etal 2014). For more information on the Magellanic System
and especially the Magellanic Clouds morphology see Introduction in
Jacyszyn-Dobrzeniecka \etal (2016) (hereafter Paper I). Here we
concentrate on an analysis based on the RR~Lyr (RRL) type variable
stars.

The RRL stars are pulsating stars of great importance. They obey the
period--luminosity law, which together with their well established
luminosities, makes them good standard candles and allows for precise
distance determinations to globular clusters and nearby galaxies. The
RRL stars represent the old population and due to their large numbers
in most stellar systems, they serve as tracers of the
three-dimensional structure, metallicity distribution, and star
formation history of galaxies. There was a great number of studies
that analyzed the Magellanic Clouds' morphology with RRL variables,
and we will summarize their main results below. All studies were based
on the third part of the Optical Gravitational Lensing Experiment
(OGLE) Catalog of Variable Stars (OCVS) containing over 17\,000 RRL
type ab (RRab) stars in the LMC (Soszyñski \etal 2009) and almost 2000
RRab stars in the SMC (Soszyñski \etal 2010). However that dataset did
not cover the extended area around the Magellanic Clouds, in contrary
to the OGLE-IV data that we use here.

The RRL stars distribution in the LMC is known to be roughly regular,
and has been often modeled as a triaxial ellipsoid (Pejcha and Stanek
2009, Deb and Singh 2014), which is rotated such that the eastern side
of this galaxy is closer to us than the western side (Pejcha and
Stanek 2009, Haschke \etal 2012a). Some studies suggested that the RRL
population of the LMC has two components: the disk and the halo
(Subramaniam and Subramanian 2009, Deb and Singh 2014), although the
existence of the disk has been questioned (Wagner-Kaiser and
Sarajedini 2013). It was also proposed, that the LMC has a bar-like
structure in the center which stands out as a RRL stars overdensity
(Subramaniam and Subramanian 2009), and is almost 5~kpc in front of
the main body of the LMC disk (Haschke \etal 2012a).

The RRL stars distribution in the SMC also has a regular, ellipsoidal
shape (Haschke \etal 2012b) that can be modeled as a triaxial
ellipsoid extended along the line-of-sight (Subramanian and
Subramaniam 2012, Deb \etal 2015). The central part of the SMC was
found to have a large line-of-sight depth (Haschke \etal 2012b), which
has been interpreted as a bulge (Deb \etal 2015). The north-eastern
side of the RRL stars distribution seems to have a larger depth
(Kapakos \etal 2010). It is also closer to us than the main SMC body
(Subramanian and Subramaniam 2012, Deb \etal 2015) and contains more
metal-rich stars (Deb \etal 2015). A study by Kapakos \etal (2011) and
Kapakos and Hatzidimitrou (2012) showed that stars with different
metallicities seem to belong to different dynamical structures. The
metal-rich objects constitute a thick disk with a bulge, while the
metal-poor stars form a halo.

In the area between the Magellanic Clouds -- the Magellanic Bridge --
intermediate age stars were observed by N\"oel \etal (2013,
2015). Moreover, candidates for an old stellar population were found
by Bagheri \etal (2013). They used 2MASS and WISE near-infrared
catalogs and found RGB and AGB stars in an on-sky stripe between the
Clouds. Authors were unable to identify whether these objects are
genuine Bridge members or they belong to the LMC or SMC halo.

Soszyñski \etal (2016ab) recently released the newest part of the OGLE
Collection of RRL stars that enabled us to analyze the
three-dimensional morphology of the Magellanic System that we present
here. The Collection is based on the OGLE-IV data (Udalski \etal 2015)
that cover about 650 square degrees in this region. This area is
significantly greater than that of the OGLE-III survey, on which the
studies described above were based. The extended coverage of the
OGLE-IV Collection includes the outskirts of the Magellanic Clouds and
the Magellanic Bridge. This allows us to deduce the actual shape of
these galaxies although the farthest outskirts, especially in the LMC
area, are still not entirely covered by observations.

We organized the paper as follows. Section~2 gives description of the
OGLE-IV data and OGLE Collection of RRL stars. In Section~3, the
technical details of the analysis are presented. We then describe the
three-dimensional structure of the LMC, SMC and Magellanic Bridge in
Sections~4, 5 and 6, respectively. Section~7 presents comparison of
the RRL stars and CCs distribution from Paper I. We summarize our
results in Section~8.

\Section{Data}
\subsection{The OGLE Collection of RR~Lyr Stars}
The newest part of the OGLE Collection of RRL stars (Soszyñski \etal
2016ab) contains 45\,453 objects in total and is the largest published
catalog of RRL stars up to date. The classification was based on the
period search for almost all {\it I}-band light curves in the OGLE
database (Udalski \etal 2015). Then light curves with periods from 0.2
to 1~day were selected and automatic and manual classification was
performed. Finally, each light curve was inspected visually. When the
case was doubtful other parameters, like the position of the object in
the color--magnitude diagram, were taken into account. About 40\% of
the RRL stars were not included in the previous versions of the OGLE
Collection of RRL stars. Almost all of them are located in the
extended region covered by OGLE-IV that was not observed during
earlier phases of the OGLE project.

The Collection includes 32\,581 RRab, 10\,246 RRc, and 2624 RRd stars,
with 22 anomalous RRd stars. Of those 39\,082 are located in the LMC,
whereas 6369 -- in the SMC. The boundary between these galaxies was
set at ${\rm RA=2\uph48\upm}$ because of a local minimum of the number
of RRL stars. This value is only an approximation because it is not
possible to separate the Magellanic Clouds due to their overlapping
halos. Similarly, it is not possible to entirely separate the
Magellanic Clouds' RRL stars from Milky Way halo's RRL stars so the
sample possibly contains some number of the latter ones. The
completeness of the OGLE Collection of RRL stars is about 96\%. The
gaps between CCD chips in the OGLE-IV camera are responsible for the
loss of about 7\% of stars from the fields that were not covered by
the OGLE-III.

\subsection{The Sample Selection}
Our analysis is based on RRL stars pulsating in the fundamental mode
(RRab). Among 32\,581 RRab stars 27\,620 are located in the LMC and
4961 in the SMC. Hereafter when we write about our RRL stars sample we
mean these RRab stars. We applied the same cuts to our sample as
described in Skowron \etal (2016). We rejected the objects that did
not have the {\it V}-band magnitude because these stars were useless
for the Wesenheit magnitude calculations. Then we removed RRL stars
with large uncertainties of phase parameters that were later used to
calculate photometric metallicities. In the next step, 20\% of objects
with the largest scatter of the light curve around the Fourier
decomposition were excluded from the sample.

\begin{figure}[b]
\centerline{\includegraphics[width=8.5cm]{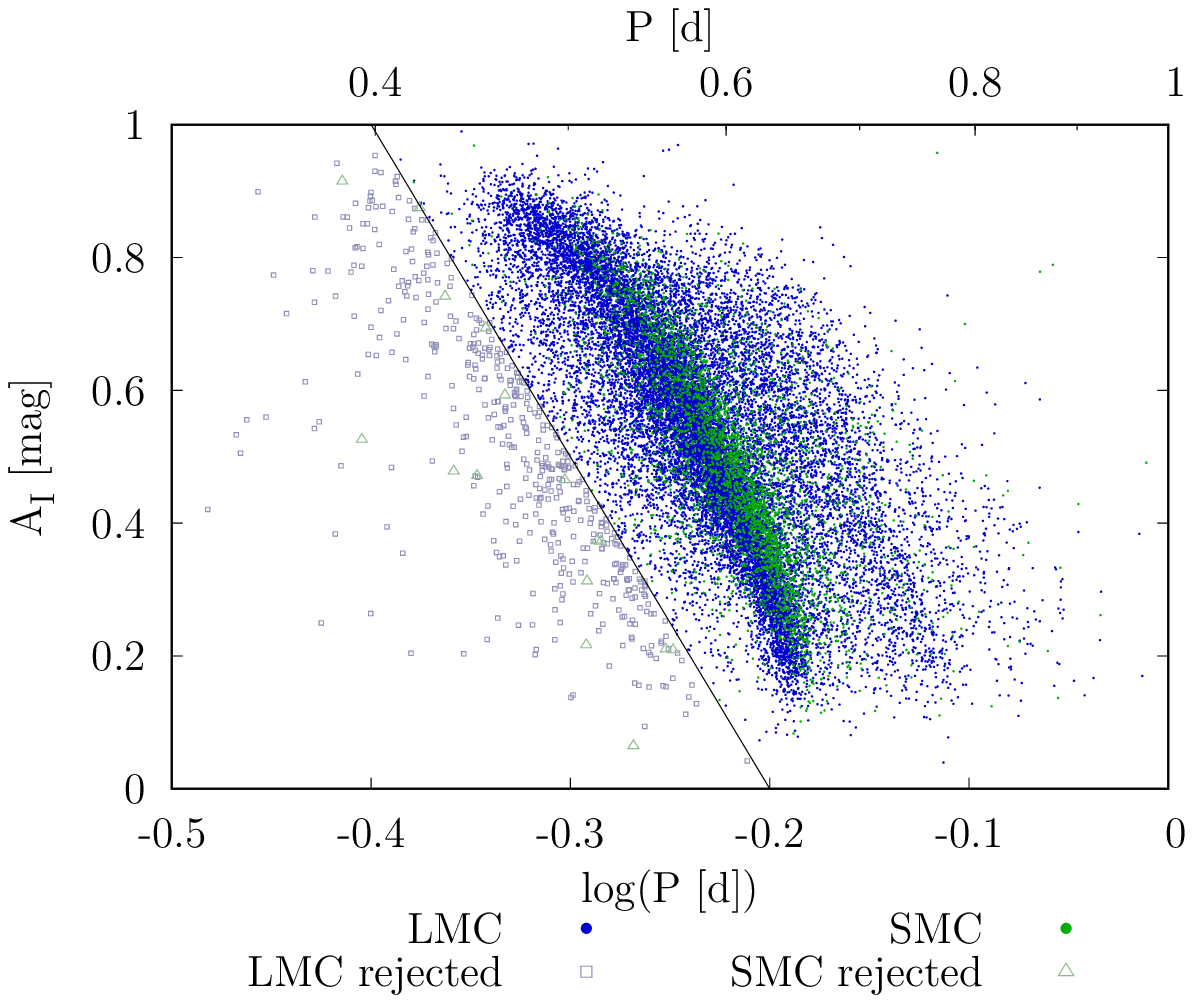}}
\FigCap{The Bailey diagram for RRL stars (ab). Black line denotes the 
adopted limit for a blend rejection. The SMC RRL stars are overplotted
on the LMC RRL stars. The rejected SMC RRL stars are marked with large
triangles while the rejected LMC RRL stars -- with squares.}
\end{figure}
After this procedure we were left with 20\,573 RRL stars in the LMC
and 3560 in the SMC. Next, we made a cut on the Bailey diagram in
order to better eliminate blends from our sample and excluded stars
with peak-to-peak amplitude lower than for a typical RRL stars at a
given period $P$ in the {\it I}-band, \ie we removed objects for which
$A_I<-5\cdot\log(P)-1$ (see Fig.~1). Then we fitted the
period--luminosity relation (P-L) to our sample and iteratively
removed RRL stars with luminosities deviating more than $3\sigma$ from
the fit (see Fig.~2). The results are described in Section~2.3. This
left us with the final sample consisting of 19\,401 RRab stars in the
LMC and 3458 stars in the SMC.
\begin{figure}[htb]
\centerline{\includegraphics[width=11.3cm]{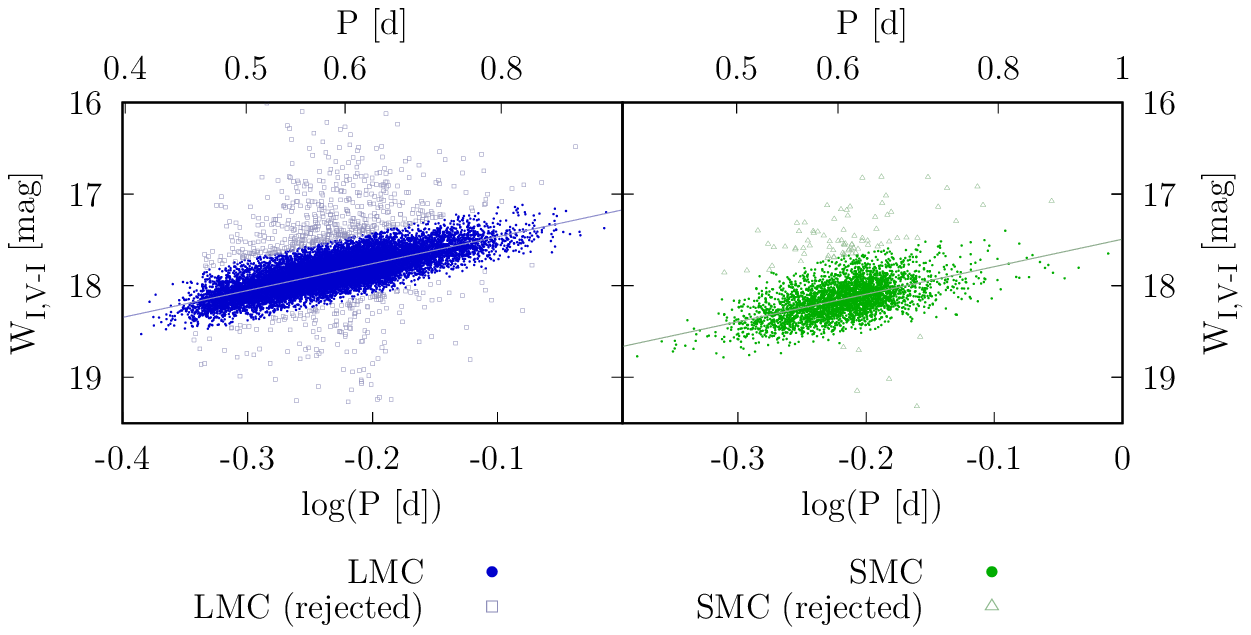}}
\FigCap{P-L relations for the Wesenheit magnitude for RRL(ab)
stars in the Magellanic System showing objects rejected as $3\sigma$
outliers during the fitting procedure. {\it Left panel:} The fit for
the LMC. Rejected objects are marked with squares. {\it Right panel:}
The fit for the SMC. Rejected objects are marked with triangles.}
\end{figure}

\begin{figure}[p]
\centerline{\includegraphics[width=12cm]{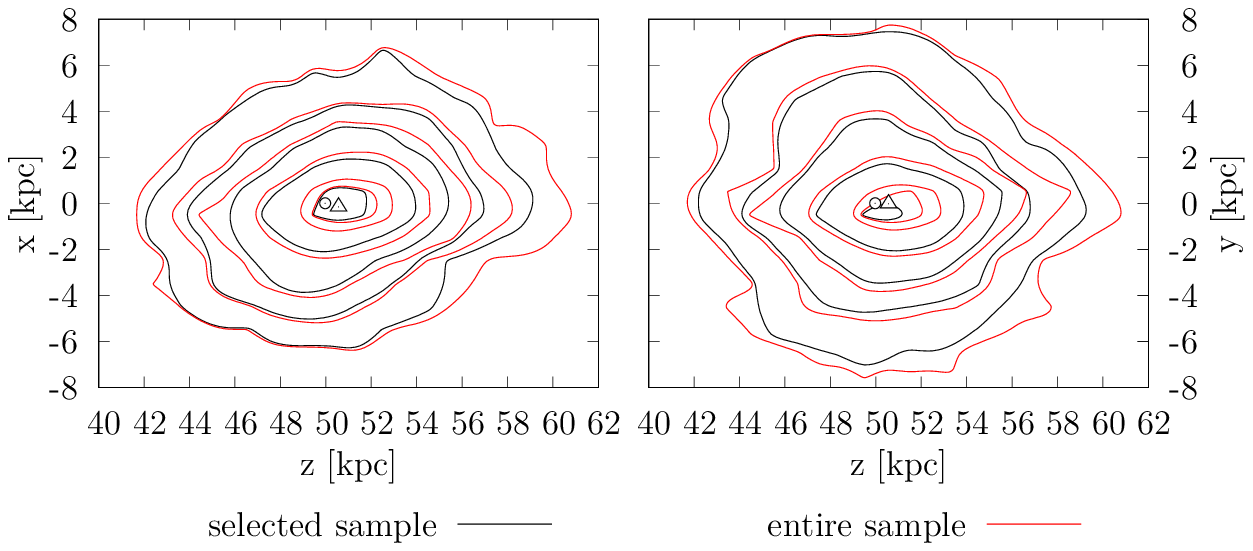}} \FigCap{Stellar
density contours of the LMC RRL stars for the entire RRL sample -- red
(objects lacking {\it I}- or {\it V}-band magnitude are not included
in this plot) and the cleaned sample -- black, on the $xz$ and $yz$
planes in the Cartesian projection. Contour levels are the same in
{\it both panels}.}
\vskip9mm
\includegraphics{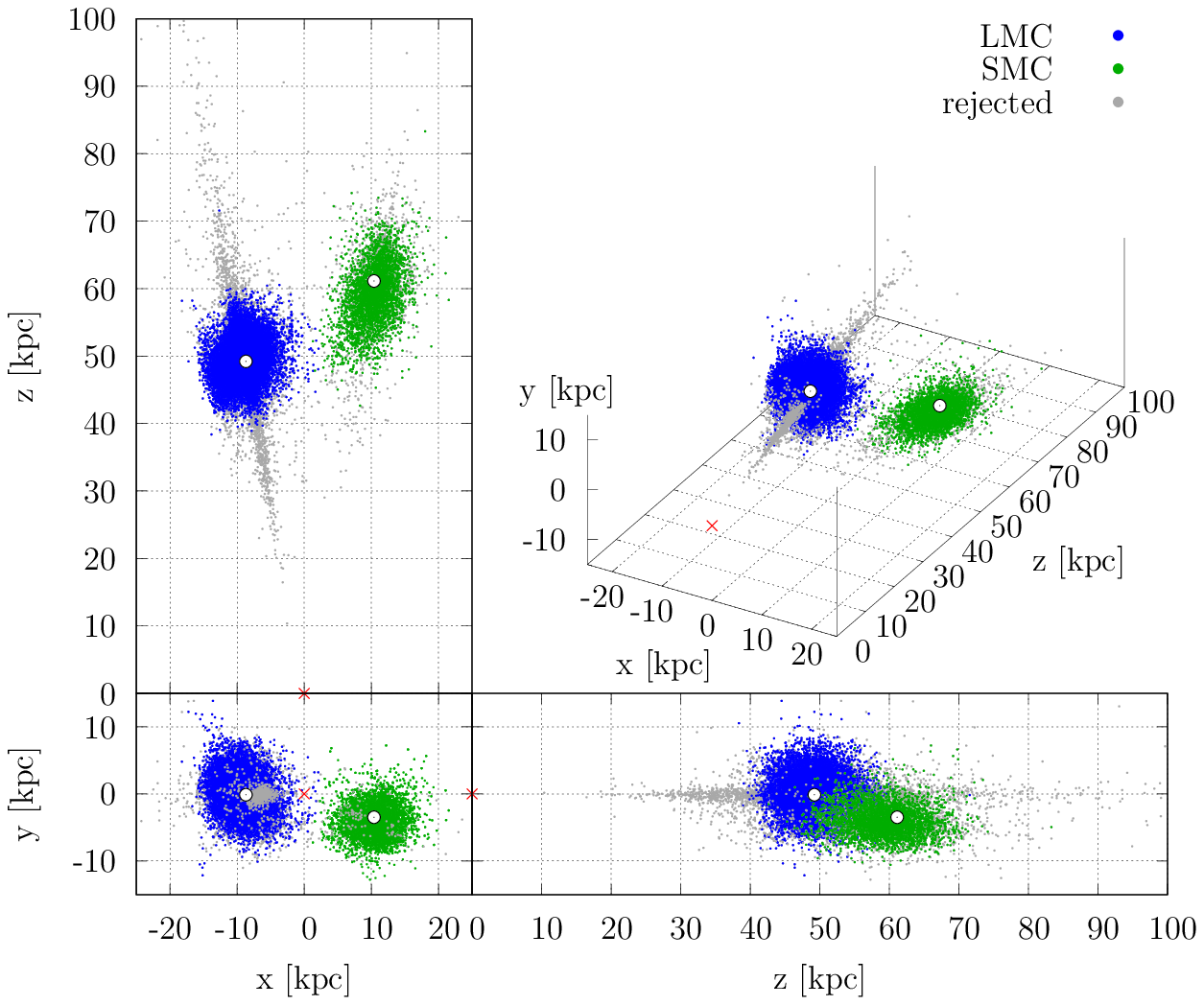}
\FigCap{The RRL stars in the Magellanic System in the Cartesian 
coordinates. The LMC stars are marked with blue dots, while the SMC
stars -- with green dots. Additionally, all the rejected RRL stars are
shown with gray color (the RRL stars lacking {\it I}- or {\it V}-band
magnitude are not included in this plot). Red cross marks the location
of the observer. White circle denotes the LMC (Pietrzyñski \etal 2013,
van der Marel and Kallivayalil 2014) and SMC (Stanimiroviæ \etal 2004,
Graczyk \etal 2014) dynamical centers on this and the following maps.}
\end{figure}
After all these restrictive cuts we expected that we would see no
blends in our data. Unfortunately, three-dimensional maps of the LMC
still show a non-physical feature -- an elongation in the LMC
structure along the line-of-sight coming out of the center of this
galaxy and visible on its both sides (hereafter we refer to it as the
LMC blend-artifact). Fig.~3 illustrates this effect on the $xz$ and
the $yz$ planes in the Cartesian projection, that will be described in
Section~3.3. Red contours represent all RRL stars, before any sample
cuts were done, while black contours show the cleaned, final
sample. The elongated central structure has decreased, but not
vanished entirely. Its cone-like shape and orientation exactly toward
the observer at (0,0,0) indisputably point to its non-physical nature.
The LMC blend-artifact is also well visible on the $xz$ plane in
Fig.~4. Unfortunately, it is very difficult to separate all the blends
from unblended stars because these objects are mixed together in every
parameter space. We tried to make additional and more restrictive cuts
on diagrams including color, magnitude, amplitude, period, but none of
these made a significant difference and the non-physical feature
remained. Instead, normal, unblended RRL stars were removed. For this
reason we refrain from performing additional cuts as this can falsify
the two-dimensional maps and distributions and lead to a lower than
real RRL stars column density. The existence of the LMC blend-artifact
requires that any analysis of the LMC center adopts a very careful
approach to the sample selection and analysis processes.

\Section{Data Analysis}
\subsection{Period--Luminosity Relation}
After removing objects with $A_I<-5\cdot\log(P)-1$ on the Bailey
diagram we fitted a period--luminosity relation to our sample. We used
the reddening-independent Wesenheit index (Madore 1976) for the {\it
V}- and {\it I}-band photometry:
$$W_{I,V-I}=I-1.55\cdot(V-I)\eqno(1)$$ 
The value of the coefficient (1.55) was calculated based on the
dependence of the {\it I}-band extinction on $E(V-I)$ reddening
(Schlegel \etal 1998). We used the least-squares method to fit the
linear function in the form:
$$W_{I,V-I}=a\cdot\log(P)+b\eqno(2)$$
separately to the LMC and SMC sample. In each iteration we rejected
RRL stars that were $3\sigma$ outliers until there were none. The
rejected objects are mostly blends, additionally affected by
crowding. The results for the Wesenheit magnitude as well as for the
{\it I}- and {\it V}-band magnitudes are shown in Table~1. Fig.~2
shows the fit for the Wesenheit magnitude and the rejected stars.

\renewcommand{\arraystretch}{1.55}
\MakeTable{c|c|c|c|r@{.}l|c|c}{12cm}{P-L relations for RRab stars in the Magellanic Clouds}
{
\hline
\multicolumn{3}{l}{P-L for Wesenheit magnitude} & \multicolumn{4}{c}{$W_{I,V-I}=a\cdot\log(P)+b$} & \\ 
\hline
Galaxy & $a$ & $b$ [mag] & $\sigma$ [mag] & \multicolumn{2}{c}{$\chi^2/{\rm dof}$} & $N_{\rm inc}$ & $N_{\rm rej}$ \\ 
\hline
LMC & $-2.933\pm0.009$ & $17.172\pm0.003$ & 0.114 & 3&605 & 19 401 & 720 \\ 
SMC & $-3.001\pm0.028$ & $17.492\pm0.007$ & 0.158 & 6&980 & 3 458 & 86 \\ 
\hline
\multicolumn{3}{l}{P-L for {\it I}-band magnitude} & \multicolumn{4}{c}{$I=a\cdot\log(P)+b$} & \\ 
\hline
Galaxy & $a$ & $b$ [mag] & $\sigma$ [mag] & \multicolumn{2}{c}{$\chi^2/{\rm dof}$} & $N_{\rm inc}$ & $N_{\rm rej}$ \\ 
\hline
LMC & $-1.680\pm0.009$ & $18.374\pm0.003$ & 0.142 & 5&587 & 19 704 & 417 \\ 
SMC & $-1.709\pm0.028$ & $18.673\pm0.007$ & 0.153 & 6&557 & 3 482 & 62 \\ 
\hline
\multicolumn{3}{l}{P-L for {\it V}-band magnitude} & \multicolumn{4}{c}{$V=a\cdot\log(P)+b$} & \\ 
\hline
Galaxy & $a$ & $b$ [mag] & $\sigma$ [mag] & \multicolumn{2}{c}{$\chi^2/{\rm dof}$} & $N_{\rm inc}$ & $N_{\rm rej}$ \\ 
\hline
LMC & $-0.910\pm0.009$ & $19.139\pm0.003$ & 0.187 & 9&768 & 19 625 & 496 \\ 
SMC & $-0.934\pm0.028$ & $19.422\pm0.007$ & 0.167 & 7&786 & 3 475 & 69 \\ 
\hline
\multicolumn{8}{p{12cm}}{$N_{\rm inc}$ is the number of objects included in the final fit, 
while $N_{\rm rej}$ is the number of rejected objects.}}

\subsection{Metallicities and Distances}
The photometric metallicity of each RRL star in our sample was calculated
the same way as in Skowron \etal (2016). The $\varphi^I_{31}$ phase
parameter from the Fourier decomposition of the {\it I}-band light curve
was transfromed to the phase parameter in the Kepler band $\varphi^{\rm
Kp}_{31}$ and then the photometric metallicity relation of Nemec \etal
(2013) was applied. For more details on the metallicity calculation see
Section~5 in Skowron \etal (2016). To calculate the distance we first
transformed the metallicity from Jurcsik (1995) scale to the Carretta \etal
(2009) scale using the relation from Kapakos \etal (2011):
$$[{\rm Fe}/{\rm H}]_C=1.001\cdot [{\rm Fe}/{\rm H}]_J-0.112.\eqno(3)$$
Then we used the coefficients from Table~5 in Braga \etal (2015) to calculate 
the absolute Wesenheit magnitude of each RRL star:
$$W_{I,V-I,abs}=a_W+b_W\cdot\log(P)+c_W([{\rm Fe}/{\rm H}]_C+0.04)\eqno(4)$$
where $a_W=-1.039\pm0.007$, $b_W=-2.524\pm0.021$ and $c_W=0.147\pm0.004$.

Finally, the distance in pc is given by:
$$d=10^{(W_{I,V-I}-W_{I,V-I,abs}+5)/5}.\eqno(5)$$

The distance uncertainty includes the OGLE photometric uncertainty which is
$\sigma_{I,V}=0.02$~mag and the uncertainty of the calculated
metallicity. The median distance uncertainty for the LMC is 1.46~kpc (3\%
relative to the median distance) and for the SMC 1.78~kpc (3\% relative to
the median distance). Fig.~4 shows the RRL stars in the Magellanic System
in three dimensions. The LMC stars are marked with blue dots, while the SMC
stars -- with green dots. Additionally, all the rejected RRL stars are
shown with gray dots.

\subsection{Coordinate Transformations}
In this paper, we present our results using two types of maps. The first
one is a two-dimensional equal-area Hammer projection. The $z$ axis is
pointing toward $\alpha_{\rm cen}=3\uph20\upm$, $\delta_{\rm
cen}=-72\arcd$. For each RRL star, $x_{\rm Hammer}$ and $y_{\rm Hammer}$
coordinates are calculated from the formulae used in Paper~I. Fig.~5 shows
the Magellanic System in the Hammer projection, where the distance is
color-coded.
\begin{figure}[htb]
\includegraphics{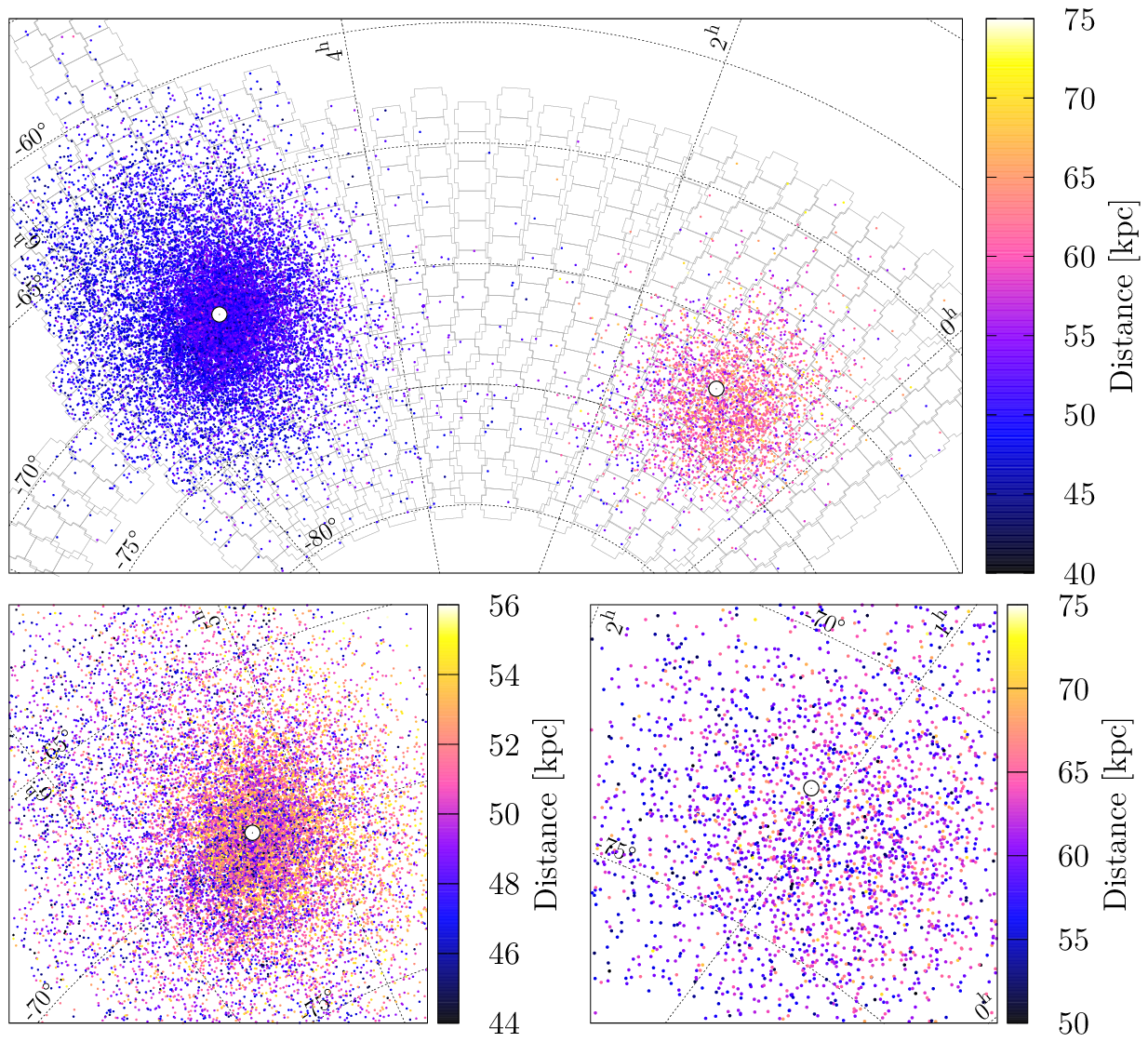}
\FigCap{The equal-area Hammer projection of the RRL stars in the 
Magellanic System with color-coded distances. Note the change in distance
range between the panels. {\it Upper panel:} The LMC is on the left while
the SMC is on the right. Gray contours represent the OGLE-IV fields. {\it
Lower left panel:} Close-up on the LMC. {\it Lower right panel:} Close-up
on the SMC (the dots representing RRL stars are one and half times larger
than on other panels). White circles mark galaxies' dynamical centers.}
\end{figure}

The second type of maps that we use shows stellar positions in the
Cartesian three-dimensional space: ($x,y,z$). We use different viewing
angles although the observer is always at $(0,0,0)$. The $z$ axis is
pointing toward different equatorial coordinates: $\alpha_{\rm cen}$ and
$\delta_{\rm cen}$. The transformation equations are the same as used in
Paper~I and were taken from van der Marel and Cioni (2001) and Weinberg and
Nikolaev (2001). Fig.~4 shows RRL stars in the Magellanic System in the
Cartesian coordinates.

Maps showing the entire Magellanic System are centered at $\alpha_{\rm
cen}=3\uph20\upm$, $\delta_{\rm cen}=-72\arcd$, while maps showing only the
LMC or SMC are centered at their dynamical centers, similarly as in
Paper~I.  For the LMC we adopted slightly different coordinates:
$\alpha_{\rm LMC-cen}=5\uph19\upm31\zdot\ups2$, $\delta_{\rm
LMC-cen}=-69\arcd35\arcm24\arcs$, which are for the whole population with a
correction for older stars proper motions (van der Marel and Kallivayalil
2014). For the SMC we use the same centering as in Paper I: $\alpha_{\rm
SMC-cen}=1\uph05\upm$, $\delta_{\rm SMC-cen}=-72\arcd25\arcm12\arcs$
(Stanimiroviæ \etal 2004). The center of each galaxy, that is marked on our
maps with a white circle, is composed of the dynamical on-sky center
($\alpha_{\rm cen}$, $\delta_{\rm cen}$) combined with the mean distance
($d$).  For the LMC we use the distance $d_{\rm LMC}=49.97\pm0.19$
(statistical) $\pm1.11$ (systematic)~kpc, calculated by Pietrzyñski \etal
(2013) which is the most accurate LMC distance up to date. For the SMC we
adopted $d_{\rm SMC}=62.1\pm1.9$~kpc from Graczyk \etal (2014). These
dynamical centers are shown in order to aid comparison with other studies
(\eg Paper~I), even though they do not comply with RRL distribution
centers.

The OGLE astrometric uncertainty is included in the Cartesian coordinates
uncertainties. This astrometric uncertainty is
$\sigma_{\alpha,\delta}=0\zdot\arcs2$. The distance uncertainty is also
included. The values of $x,y$ and $z$ position uncertainties are as
follows: $0.1~{\rm kpc}<\sigma_x<0.9~{\rm kpc}$, $0.1~{\rm kpc}<
\sigma_y<0.7~{\rm kpc}$, and $1.3~{\rm kpc}<\sigma_z<4.1~{\rm kpc}$.

The most important parameters of the RRL stars sample analyzed in this
publication are available on-line in a tabular form from the OGLE website:
\vskip5pt
\centerline{\it http://ogle.astrouw.edu.pl} 
\vskip5pt
Table~2 presents the first few lines of the file.

\setlength{\tabcolsep}{2pt}
\MakeTableee{cccccc}{12cm}{RRL stars (ab) in the Magellanic System}
{\hline
\multicolumn{6}{c}{Columns 1--6} \\ 
\hline
Location & OCVS Id & P [d] & {\it I} [mag] & {\it V} [mag] & $W_{I,V-I}$ [mag]\\ 
\hline
LMC & OGLE-LMC-RRLYR-00001 &  0.6347521 & 18.772 & 19.455 & 17.713  \\ 
LMC & OGLE-LMC-RRLYR-00003 &  0.6564971 & 18.649 & 19.306 & 17.631  \\ 
LMC & OGLE-LMC-RRLYR-00005 &  0.6433519 & 18.942 & 19.613 & 17.902  \\ 
$\vdots$ & $\vdots$ & $\vdots$ & $\vdots$ & $\vdots$ & $\vdots$ \\ 
\cline{1-6}
\multicolumn{6}{c}{Columns 7-13} \\ 
\hline
$[{\rm Fe}/{\rm H}]_N$ & RA ~~~~~~~~~~~~~~~~~ Dec & d [kpc] & $x^{(a)}$ [kpc] & $y^{(a)}$\ [kpc] & $z^{(a)}$ [kpc] \\ \hline
$-1.63 \pm 0.12$ & $04\uph27\upm45\zdot\ups45$ ~ $-70\arcd43\arcm12\zdot\arcs0$ &  $50.23 \pm 1.46$ &  $-4.83 \pm 0.39$ &   $0.44 \pm 0.85$ &  $49.99 \pm 1.54$ \\
$-1.41 \pm 0.11$ & $04\uph28\upm08\zdot\ups50$ ~ $-70\arcd21\arcm22\zdot\arcs8$ &  $48.44 \pm 1.39$ &  $-4.77 \pm 0.38$ &   $0.71 \pm 0.82$ &  $48.20 \pm 1.48$ \\
$-1.14 \pm 0.42$ & $04\uph28\upm21\zdot\ups06$ ~ $-70\arcd08\arcm54\zdot\arcs5$ &  $53.33 \pm 2.13$ &  $-5.32 \pm 0.45$ &   $0.96 \pm 0.90$ &  $53.06 \pm 2.19$ \\
$\vdots$ & $\vdots$ & $\vdots$ & $\vdots$ & $\vdots$ & $\vdots$ \\ 
\cline{1-6}
\multicolumn{6}{p{12cm}}{The electronic version of the entire 
sample used in this study is available on-line from the OGLE website. 
{\it (a)} The Cartesian $x,y$, and $z$ coordinates.}
}

\subsection{Model and Ellipsoid Fitting}
In the next step, we modeled the RRL stars spatial distribution by fitting
triaxial ellipsoids to surfaces of a constant number density, to the LMC
and SMC three-dimensional data in the Cartesian coordinate space. First, we
calculated the local density of RRL stars in a $2\times2\times 2$~kpc cube
around each star, which was up to 338 and 29 stars per~kpc$^3$ in the LMC
and SMC, respectively. The cube size was chosen as a trade-off between the
resolution and smoothness of the resulting star density distribution. 
Subsequently, we divided both samples of RRL stars into bins of nearly
constant star density and then fitted triaxial ellipsoids to these
subsamples using an algorithm proposed by Turner \etal (1999), described
below.

We aimed to find the parameters of an ellipsoid given its quadratic form:
$$ax^2+by^2+cz^2+dxy+exz+fyz+gx+hy+kz+l=0.\eqno(6)$$
We found the best-fit ellipsoid by minimizing the sum of squared distances
between the data points and the modeled ellipsoid. The resulting quadratic
forms were then transformed to obtain parameters of the ellipsoid:
coordinates of the center, length of semi-axes, and their orientation. The
uncertainties were estimated using the bootstrap method. To aid the
comparison with previous works, we provide two parameters describing the
orientation of ellipsoids: inclination and position angle of the longest
axis.

If an ellipsoid is centered at the origin, then its quadratic form is
${\rm\bf X^TAX}=C$, where $C>0$ and {\bf A} is a symmetric matrix with
positive eigenvalues, ${\rm\bf X^T}=[x,y,z]$. From the principal axis
theorem, we know that eigenvectors of a matrix {\bf A} form an orthonormal
basis such as ${\rm\bf P^TAP=D}$, where {\bf D} is a diagonal matrix
and {\bf P} is a square matrix consisting of the eigenvectors corresponding
to the eigenvalues in {\bf A}. In that basis, the quadratic form is simply
${\rm\bf X^TA X}=\sum_i\lambda_ix_i^2=C$, and hence the semi-axes of
the ellipsoid are equal to $\sqrt{C/\lambda_i}$, where $\lambda_i$ are
eigenvalues of {\bf A}. Eigenvectors of {\bf A} span the semi-axes.

It can be straightforwardly shown that:
\[
{\rm \bf A} = 
\begin{bmatrix}
a & d/2 & e/2 \\
d/2 & b & f/2 \\
e/2 & f/2 & c 
\end{bmatrix}
\]
For the ellipsoid centered at ${\rm\bf X_0}$: 
$${\rm\bf (X-X_0)^TA(X-X_0)=X^TAX}-2{\rm\bf X^TAX_0+X_0^TAX_0}=C.\eqno(7)$$

Hence, the origin of the ellipsoid 
\[
{\rm\bf X_0}=-\frac{1}{2}{\rm \bf A}^{-1}\begin{bmatrix}
g \\
h \\
k
\end{bmatrix}
\]
while $C={\rm\bf X_0^TAX_0}-l$.

\Section{The Large Magellanic Cloud}
\subsection{Three-Dimensional Structure}
The RRL stars distribution in the LMC is known to be roughly regular, or
ellipsoidal, possibly with a bar (Pejcha and Stanek 2009, Subramanian and
Subramaniam 2012, Haschke \etal 2012a, Wagner-Kaiser and Sarajedini 2013,
Deb and Singh 2014).

We have estimated the sample center parameters using the maxima of the
Right Ascension, Declination and distance of the RRL stars distribution
which are $\tilde{\alpha}_{\rm LMC}{=}\break5\uph21\upm31\zdot\ups2$,
$\tilde{\delta}_{\rm LMC}=-69\arcd36\arcm36\arcs$, $\tilde{d}_{\rm
LMC}=50.56$~kpc (hereafter the distribution center). The median LMC RRL
stars distance based on our data is $d_{\rm LMC,med}=50.64$ kpc. This is
slightly different than the dynamical center coordinates derived by van der
Marel and Kallivayalil (2014) which were $\alpha_{\rm
LMC-cen}=5\uph19\upm31\zdot\ups2$, $\delta_{\rm LMC-cen}=
-69\arcd35\arcm24\arcs$ and the mean LMC distance from Pietrzyñski \etal
(2013) derived from eclipsing binaries: $d_{\rm LMC}=49.97\pm0.19$
(statistical) $\pm1.11$ (systematic)~kpc.

Fig.~4 shows the Magellanic System in the Cartesian coordinates where the
LMC reveals its regular, although not entirely symmetrical, shape in three
dimensions. The most protruding ``substructure'' is the LMC blend-artifact
-- a non-physical structure build up of the RRL stars seemingly drawn-out
of the galaxy along the line-of-sight. These stars are mostly blends,
additionally affected by crowding effects and are located in the dense LMC
center. Because of their relatively low luminosity, RRL stars are very
prone to such blending and crowding effects. As we have already described
in Section~2.2 it is impossible to remove all the blends from our sample
because many of them are not distinguishable from unblended RRL stars based
solely on their light curves. An attempt to do so would lead to
non-physical results.
\begin{figure}[htb]
\centerline{\includegraphics[width=13cm]{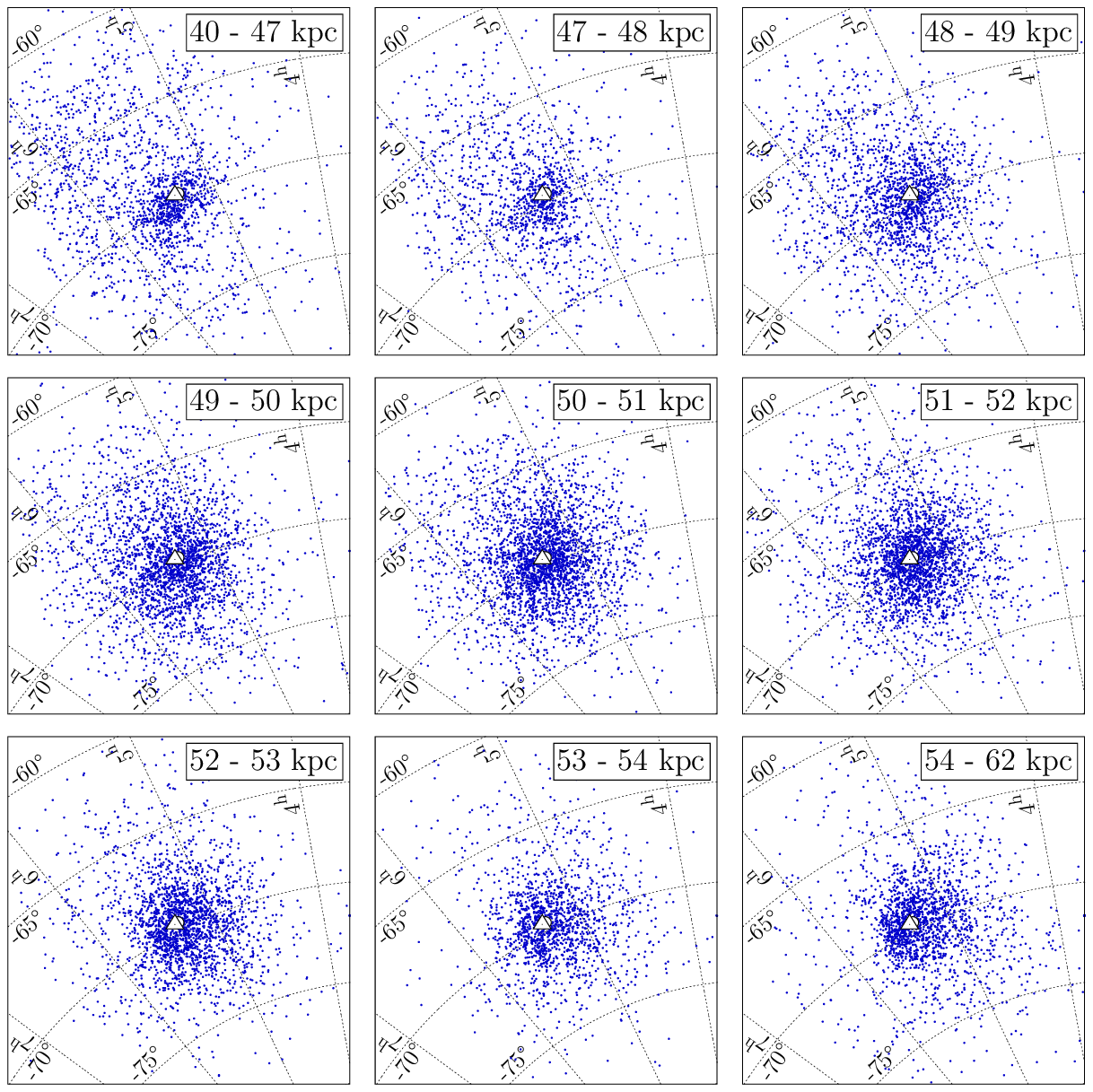}}
\FigCap{Distance tomography of the RRL stars distribution in the
LMC in the Hammer equal-area projection. Note different distance
ranges. White circle marks the LMC dynamical center. White triangle marks
the RRL stars distribution maxima along the RA, Dec and distance axes.}
\end{figure}

The on-sky projection of the LMC seems to be roughly regular (see
Fig.~5). To further investigate the three-dimensional structure of this
galaxy we show its distance tomography in Fig.~6. The upper row represents
the closest RRL stars in the LMC. There is a well visible clump at the
center, elongated in the east-west direction and concentrated more on the
eastern side of the distribution and dynamical center (first panel). It may
seem to constitute the LMC bar, similarly as in Fig.~5 from Haschke \etal
(2012a), but in fact this is a reflection of the non-physical LMC 
blend-artifact. On the other hand, we see that the LMC extended halo and
the closest parts of it are definitely concentrated in the north-eastern
parts of this galaxy. The LMC halo is not symmetrical with respect to the
distribution and the dynamical center of this galaxy.

The middle row shows RRL stars near the average LMC distance. Here, the
central parts of the LMC have a more regular shape. Again, we see that the
LMC RRL stars halo is more numerous in the north-eastern parts of this
galaxy. The lowest row represents the farthest LMC RRL stars. The RRL stars
in the central regions are more clumped on the eastern side but this is
again due to the LMC blend-artifact as it is consistent with the
distribution maximum. Interestingly, the LMC halo's farthest parts are more
numerous on the western side. This is the direction toward the SMC. The
distance tomography of the LMC suggests that the eastern part of the LMC is
closer than the western part.

Column density maps in three Cartesian dimensions are shown in Fig.~7. The
bin size is 0.5~kpc along each axis. On the $xz$ and $yz$ planes the LMC
blend-artifact is clearly visible as a longitudinal structure that is
elongated along the line-of-sight. The ``view from the top'' -- $xz$ plane
-- shows that the LMC outskirts are asymmetrical with the eastern side
located closer to us than the western side.  The LMC halo seems to be
neither spheroidal nor ellipsoidal, which is also prominent on the $yz$
plane.
\begin{figure}[htb]
\centerline{\includegraphics[width=13cm]{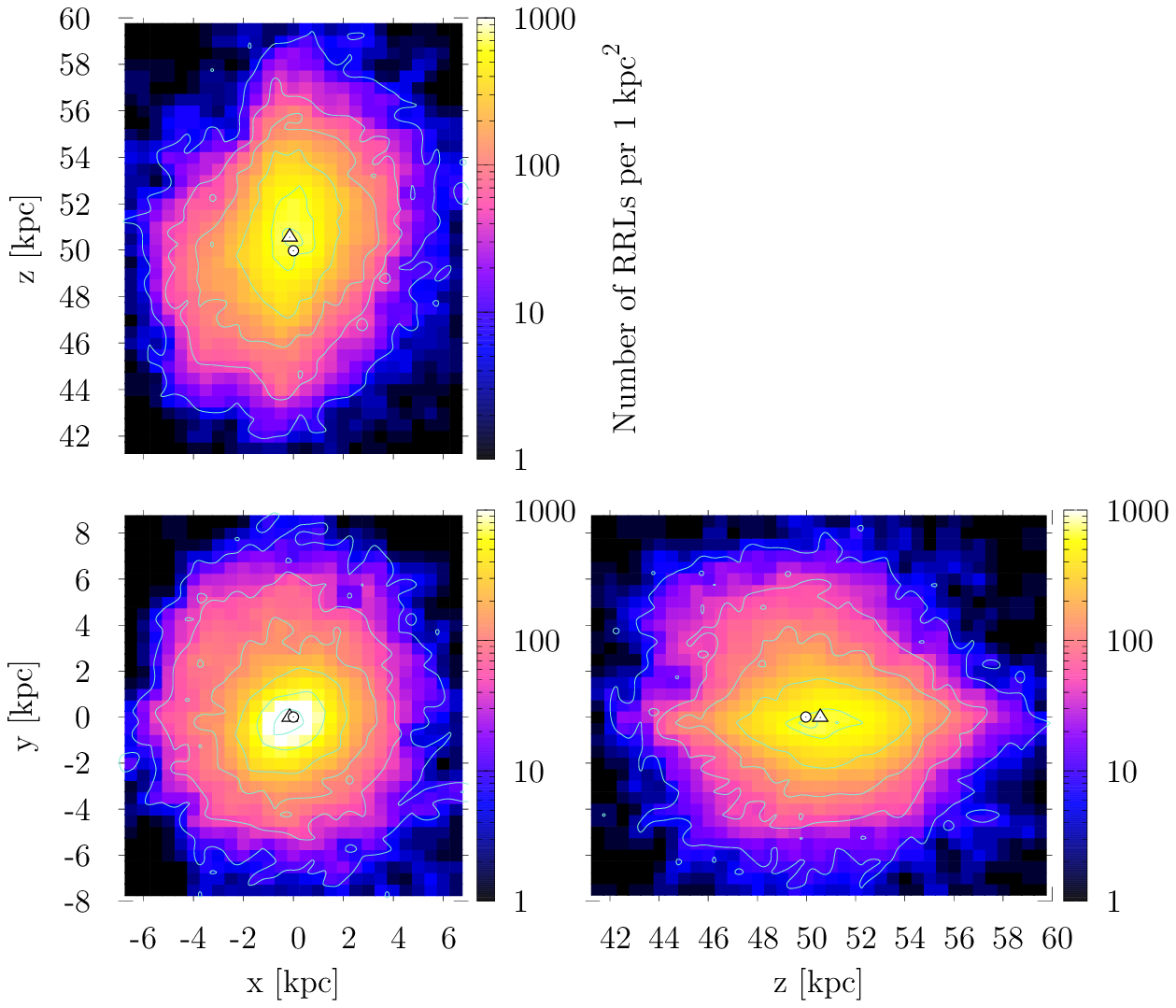}}
\vskip5pt
\FigCap{RRL stars density maps in the LMC in the Cartesian coordinates 
(the $z$ axis is pointing toward the LMC dynamical center). Bin size is
0.5~kpc in $x$, $y$ and $z$ axis. Contour levels on the $xy$ plane are 10,
50, 120, 300, 700, 1300, on the $xz$ and $yz$ 10, 50, 120, 300, 600, 700
RRL stars per 1~kpc$^2$. Note the LMC blend-artifact clearly visible on the
$xz$ and $yz$ planes. White circle and triangle mark the LMC dynamical and
distribution centers, respectively.}
\end{figure}

\subsection{Ellipsoid Fitting}
As a result of the analysis based on the two- and three-dimensional maps we
decided to model the LMC RRL stars distribution as a triaxial
ellipsoid. The LMC RRL stars were divided into 21 subsamples consisting of
135 to 963 objects. The technical details of the modeling procedure were
described in Section~3.4. The fitting results are presented in Figs.~8, 9,
and 10 and in Table~3. To minimize the influence of the non-physical LMC
blend-artifact, we decided to exclude the central region of the LMC from
the fit and the following analysis. We removed RRL stars located within an
angular on-sky radius of $1\zdot\arcd5$ from the LMC distribution
center, \ie all RRL stars along the line-of-sight in a cone (see Fig.~9).
\begin{figure}[htb]
\centerline{\includegraphics[width=12.5cm]{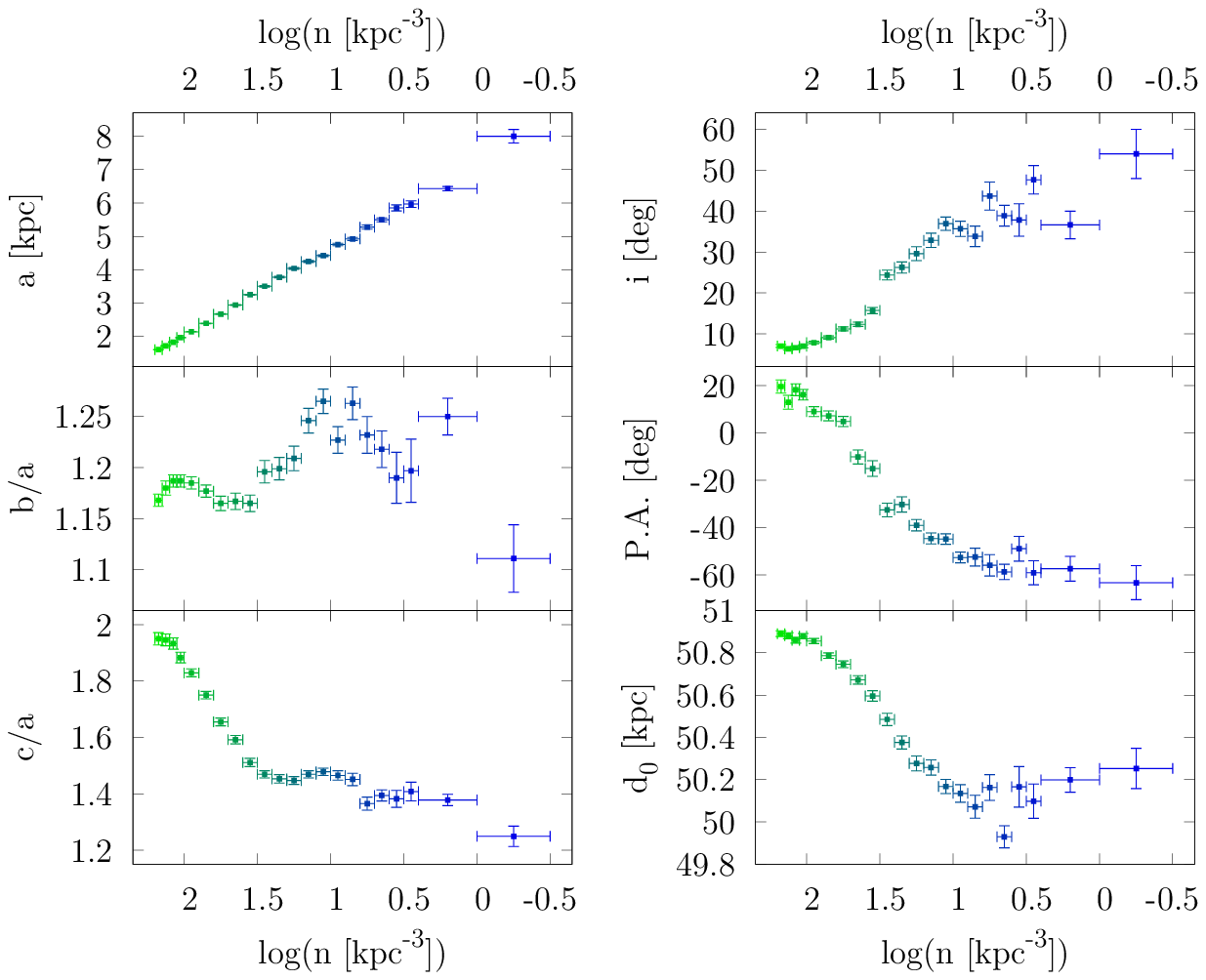}}
\vskip3pt
\FigCap{Parameters of the best-fit triaxial ellipsoids for the LMC RRL
stars. We excluded objects located within an angular radius of
$1\zdot\arcd5$ from the LMC center because of the LMC blend-artifact. Green
points represent the innermost ellipsoids while blue points -- the
outermost.}
\end{figure}

The innermost LMC ellipsoid corresponds to the star density of
$\log(n)=2.15{-}\break 2.2~{\rm kpc}^{-3}$. The axis ratio $a:b:c$ is
$1:1.168:1.950$ and it is the ellipsoid with the highest $c/a$ ratio. The
inclination is relatively small ($i=7\zdot\arcd03$), while the position
angle is large (${\rm P.A.}=19\zdot\arcd57$). As the number density $n$
decreases (\ie $a$ increases), $c/a$ ratios are decreasing while $b/a$
ratios do not show any trend (see Table~3 and Fig.~8). This shows that the
innermost region of the LMC has the most elongated shape. This effect may
not be entirely physical due to the residual blends which may cause the
central ellipsoids to be more elongated along the line-of-sight. It is not
possible to state how big this effect is, and whether it is entirely due to
the crowding and blending effects, or the inner parts of the LMC are truly
elongated as shown in the plots.

The largest ellipsoid has axis ratio $1:1.250:1.378$. We intentionally
chose $\log(n)=0.0-0.4~{\rm kpc}^{-3}$ as the largest ellipsoid because
$\log(n)=-0.5-0.0~{\rm kpc}^{-3}$ stretches farther than the OGLE-IV fields
and may not represent physical results. With increasing $a$, $i$ is also
increasing, but P.A. is decreasing (see Fig.~8). For ${\rm log}(n)=0.0-0.4\
{\rm kpc}^{-3}$: $i=36\zdot\arcd61$ and ${\rm P.A.}=-57\zdot\arcd32$. The
largest ellipsoids are less stretched, their longest axes are more inclined
and rotated differently. The median axis ratio is $1:1.23:1.45$.

\begin{figure}[tb]
\centerline{\includegraphics[width=13cm]{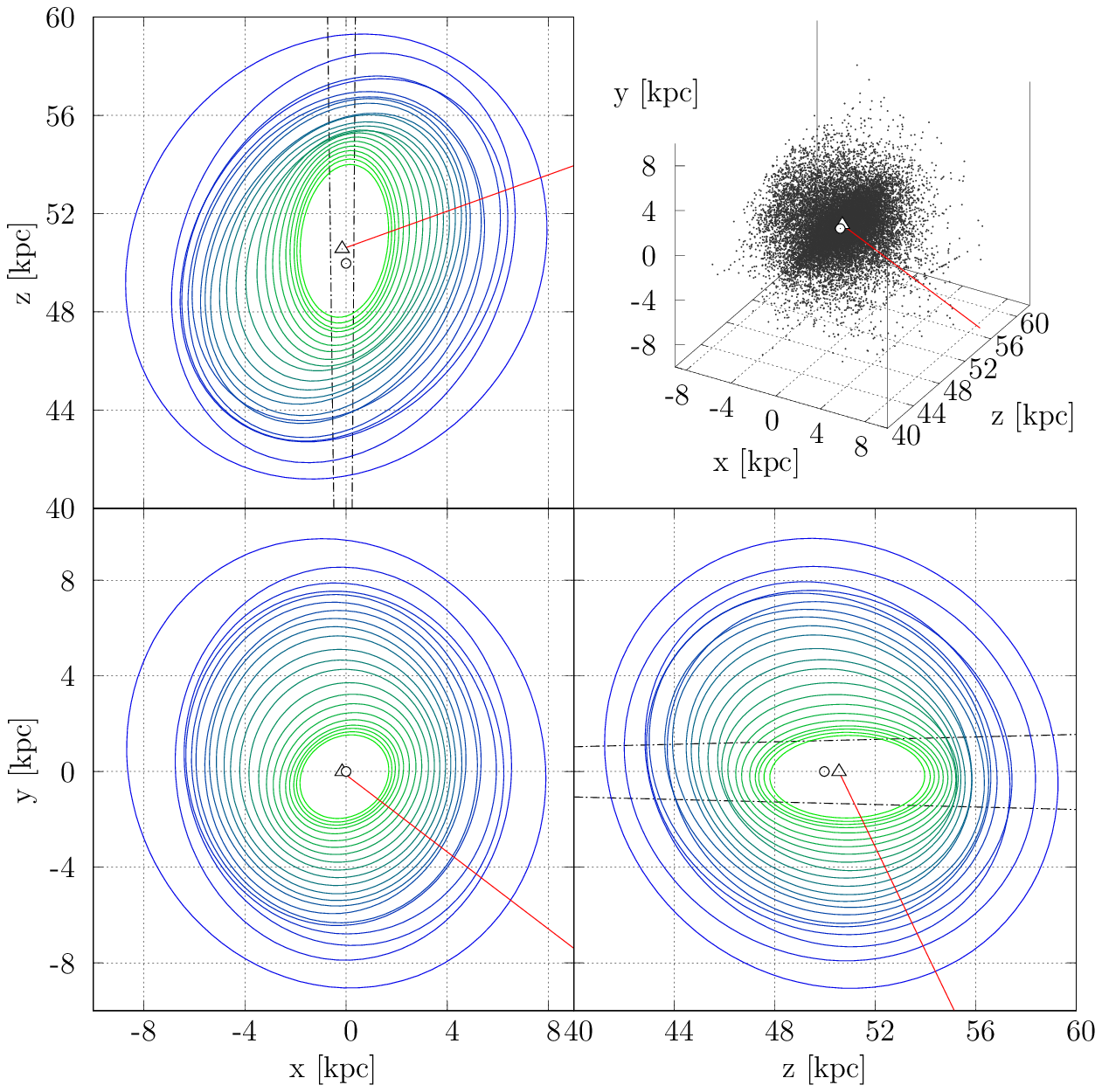}}
\vskip5pt
\FigCap{Best-fit triaxial ellipsoids for the LMC data. Dash-dotted
lines on the $xz$ and $yz$ planes represent area from where the RRL stars were 
excluded ($1\zdot\arcd5$ angular radius from the LMC distribution center). 
Colors are compatible with Fig.~8. White circle and triangle mark the LMC 
dynamical and distribution centers, respectively. Red line connects the LMC and 
SMC distribution centers.}
\end{figure}

Fig.~9 shows projections of the ellipsoids in the Cartesian space. Red
line connects the LMC and SMC distribution centers. Larger ellipsoids
do not evidently twist toward the SMC although the increasing
P.A. suggests so. On the other hand, the $xz$ and $yz$ projections
demonstrate that the LMC halo is stretched toward its smaller neighbor
more than the inner parts.

The last three columns of Table~3 represent Right Ascension,
Declination and distance of the ellipsoids' centers. We have
additionally presented the Cartesian space projections of those
centers in Fig.~10. Red line connects the LMC and SMC centers while
the black line denotes the LMC -- Milky Way centers connection. Green
points stand for the smallest ellipsoids, while blue points for the
largest. From Fig.~10 it is clearly visible that with increasing RRL
stars number the center moves farther away from the SMC -- in the
opposite direction. This is consistent with our conclusions from
Section~4.3. The LMC's farthest parts are more numerous in
north-eastern parts of this galaxy.

\begin{landscape}
\setlength{\tabcolsep}{5pt}
\renewcommand{\arraystretch}{1.15}
\MakeTableee{|c|cccccccc|}{12cm}{Triaxial ellipsoid best-fit parameters for the LMC}
{
\hline
\douprule
$\log\left(n~\left[{\rm kpc}^{-3}\right]\right)$ & $a$ [kpc] & $b/a$ & $c/a$ & $i$ [deg] & P.A. [deg] & $\alpha_0$ [deg] & $\delta_0$ [deg] & $d_0$ [kpc] \\
\hline
$2.15{-} 2.2$  & $1.606\pm0.006$ & $1.168\pm0.006$ & $1.950\pm0.022$ & ~$7.03\pm0.37$ & $19.57\pm2.79$  & $80.100\pm0.014$ & $-69.833\pm0.005$ & $50.891\pm0.014$ \\ 
$2.1 {-} 2.15$ & $1.721\pm0.006$ & $1.180\pm0.007$ & $1.946\pm0.021$ & ~$6.30\pm0.30$ & $12.90\pm2.93$  & $80.107\pm0.015$ & $-69.842\pm0.005$ & $50.881\pm0.014$ \\ 
$2.05{-} 2.1$  & $1.833\pm0.007$ & $1.187\pm0.006$ & $1.933\pm0.020$ & ~$6.61\pm0.34$ & $18.21\pm2.51$  & $80.077\pm0.018$ & $-69.835\pm0.006$ & $50.860\pm0.014$ \\ 
$2.0 {-} 2.05$ & $1.967\pm0.007$ & $1.187\pm0.006$ & $1.883\pm0.019$ & ~$6.95\pm0.37$ & $16.15\pm2.22$  & $80.094\pm0.018$ & $-69.838\pm0.006$ & $50.879\pm0.013$ \\ 
$1.9 {-} 2.0$  & $2.143\pm0.007$ & $1.185\pm0.006$ & $1.829\pm0.014$ & ~$7.86\pm0.32$ & $8.95\pm2.08$   & $80.073\pm0.018$ & $-69.837\pm0.006$ & $50.856\pm0.012$ \\ 
$1.8 {-} 1.9$  & $2.394\pm0.008$ & $1.177\pm0.006$ & $1.750\pm0.013$ & ~$9.10\pm0.40$ & ~$7.17\pm2.08$  & $80.090\pm0.019$ & $-69.835\pm0.006$ & $50.787\pm0.013$ \\ 
$1.7 {-} 1.8$  & $2.671\pm0.011$ & $1.165\pm0.007$ & $1.655\pm0.014$ & $11.21\pm0.47$ & ~$4.82\pm2.17$  & $80.074\pm0.024$ & $-69.779\pm0.008$ & $50.746\pm0.016$ \\ 
$1.6 {-} 1.7$  & $2.941\pm0.012$ & $1.167\pm0.008$ & $1.592\pm0.015$ & $12.32\pm0.52$ & $-10.15\pm2.95$ & $80.128\pm0.026$ & $-69.724\pm0.011$ & $50.672\pm0.020$ \\ 
$1.5 {-} 1.6$  & $3.251\pm0.014$ & $1.165\pm0.008$ & $1.511\pm0.015$ & $15.69\pm0.77$ & $-15.06\pm3.32$ & $80.199\pm0.032$ & $-69.633\pm0.013$ & $50.596\pm0.025$ \\ 
$1.4 {-} 1.5$  & $3.504\pm0.021$ & $1.196\pm0.011$ & $1.469\pm0.013$ & $24.36\pm1.19$ & $-32.53\pm2.91$ & $80.447\pm0.036$ & $-69.505\pm0.016$ & $50.485\pm0.029$ \\ 
$1.3 {-} 1.4$  & $3.778\pm0.024$ & $1.199\pm0.011$ & $1.453\pm0.015$ & $26.22\pm1.34$ & $-30.21\pm3.20$ & $80.681\pm0.046$ & $-69.460\pm0.019$ & $50.376\pm0.031$ \\ 
$1.2 {-} 1.3$  & $4.041\pm0.027$ & $1.209\pm0.012$ & $1.447\pm0.014$ & $29.60\pm1.70$ & $-39.00\pm2.39$ & $80.906\pm0.046$ & $-69.393\pm0.021$ & $50.277\pm0.035$ \\ 
$1.1 {-} 1.2$  & $4.249\pm0.030$ & $1.246\pm0.012$ & $1.469\pm0.013$ & $32.86\pm1.73$ & $-44.59\pm2.32$ & $80.975\pm0.052$ & $-69.268\pm0.023$ & $50.258\pm0.036$ \\ 
$1.0 {-} 1.1$  & $4.424\pm0.027$ & $1.265\pm0.012$ & $1.478\pm0.013$ & $36.91\pm1.62$ & $-44.82\pm2.23$ & $81.159\pm0.050$ & $-69.207\pm0.024$ & $50.168\pm0.034$ \\ 
$0.9 {-} 1.0$  & $4.755\pm0.035$ & $1.227\pm0.013$ & $1.465\pm0.017$ & $35.67\pm1.85$ & $-52.58\pm2.28$ & $81.202\pm0.069$ & $-69.141\pm0.032$ & $50.135\pm0.042$ \\ 
$0.8 {-} 0.9$  & $4.921\pm0.046$ & $1.263\pm0.016$ & $1.451\pm0.022$ & $33.84\pm2.51$ & $-52.43\pm3.78$ & $81.229\pm0.090$ & $-69.131\pm0.038$ & $50.072\pm0.054$ \\ 
$0.7 {-} 0.8$  & $5.277\pm0.060$ & $1.232\pm0.018$ & $1.365\pm0.023$ & $43.65\pm3.44$ & $-55.88\pm4.50$ & $81.072\pm0.104$ & $-69.152\pm0.044$ & $50.163\pm0.061$ \\ 
$0.6 {-} 0.7$  & $5.495\pm0.059$ & $1.218\pm0.018$ & $1.394\pm0.020$ & $38.86\pm2.51$ & $-58.67\pm3.22$ & $81.116\pm0.125$ & $-69.033\pm0.039$ & $49.930\pm0.052$ \\ 
$0.5 {-} 0.6$  & $5.851\pm0.091$ & $1.190\pm0.025$ & $1.382\pm0.030$ & $37.82\pm3.99$ & $-48.89\pm5.23$ & $80.527\pm0.208$ & $-69.162\pm0.069$ & $50.166\pm0.096$ \\ 
$0.4 {-} 0.5$  & $5.967\pm0.099$ & $1.197\pm0.031$ & $1.408\pm0.033$ & $47.63\pm3.48$ & $-59.03\pm5.08$ & $80.147\pm0.250$ & $-69.243\pm0.080$ & $50.098\pm0.081$ \\ 
$0.0 {-} 0.4$  & $6.430\pm0.064$ & $1.250\pm0.018$ & $1.378\pm0.020$ & $36.61\pm3.37$ & $-57.32\pm5.21$ & $80.216\pm0.169$ & $-69.215\pm0.062$ & $50.199\pm0.058$ \\ 
$-0.5{-} 0.0^{\star}$ & $8.001\pm0.204$ & $1.111\pm0.033$ & $1.249\pm0.036$ & $53.99\pm6.00$ & $-63.25\pm7.19$ & $81.201\pm0.682$ & $-69.181\pm0.156$ & $50.253\pm0.095$ \\
\hline
\noalign{\vskip5pt}
\multicolumn{9}{p{19.7cm}}{*This ellipsoid may not represent physical results due to its size extending 
farther than the OGLE-IV sky coverage in the east.}
}
\end{landscape}

\begin{figure}[h]
\centerline{\includegraphics[width=12cm]{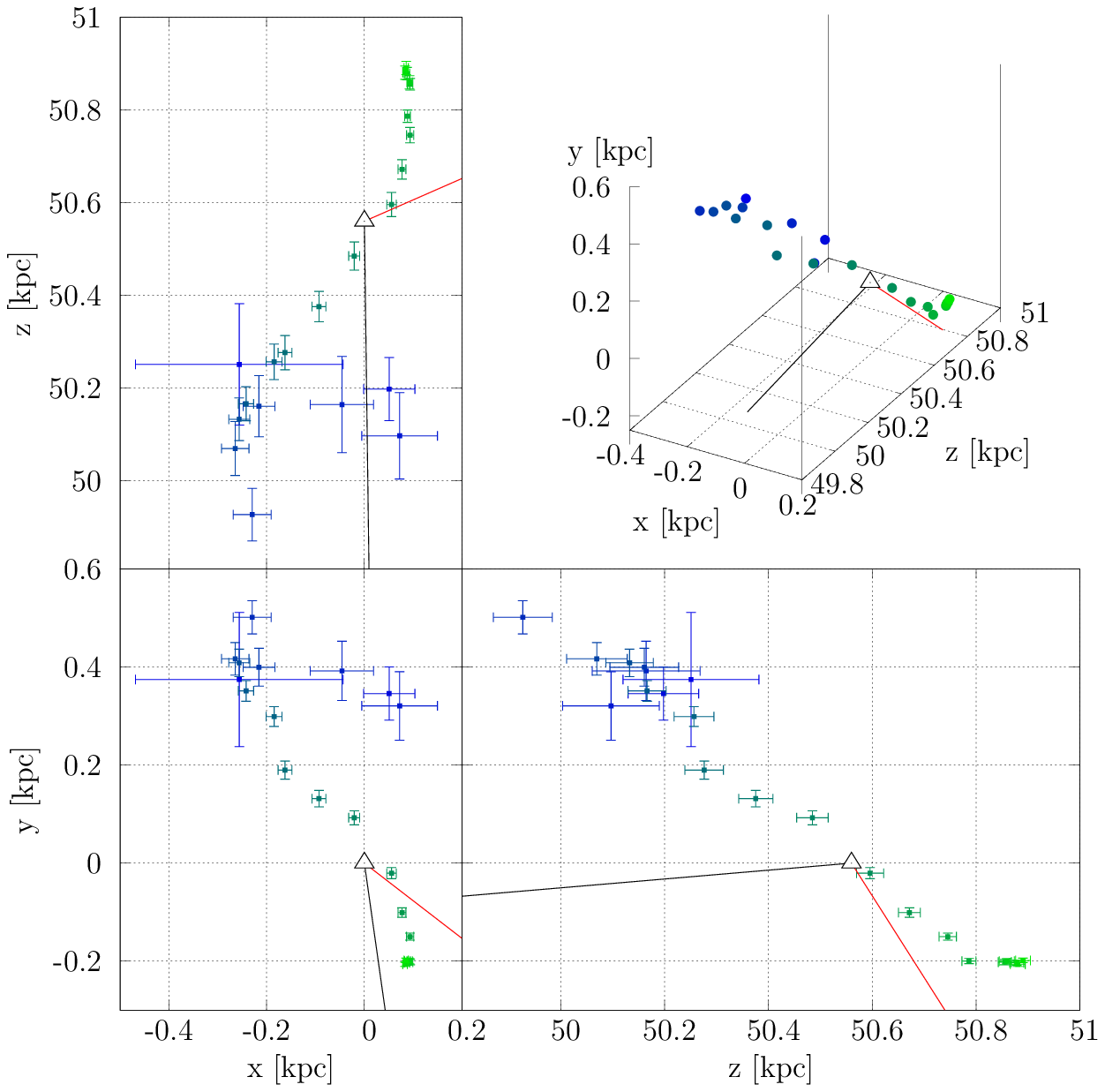}}
\FigCap{Best-fit triaxial ellipsoid centers in Cartesian coordinates 
projections for the LMC data. Colors are compatible with Figs.~8 and
9. White triangle marks the RRL stars distribution center. Red line
connects the LMC and SMC distribution centers and black line
connects the LMC distribution center with the Milky Way center
(Boehle \etal 2016).}
\end{figure}

\subsection{Comparison with Previous Studies}
Table~4 shows a comparison of RRL stars sample modeling parameters in
different studies. The $b/a$ ratio obtained from the OGLE-III data was
larger than values presented in this work even for the smallest
ellipsoids (\ie $\log(n)=2.15-2.2~{\rm kpc}^{-3}$). The closest result
to ours was presented by Pejcha and Stanek (2009) by removing RRL
stars outside 250 per square degree contour. The differences may also
be caused by the removal of stars located within the angular radius of
$1\zdot\arcd5$ from the LMC center from our sample.

The $c/a$ ratio (of the shortest to the longest ellipsoid axis) is
also smaller in our analysis, \ie our ellipsoids are less stretched,
and this difference is even more prominent. It may be due to the
restricted OGLE-III coverage or/and the LMC blend-artifact that may
distort the results. The inclination angle for larger ellipsoids

\begin{landscape}
\begin{centering}
\renewcommand{\tabcolsep}{3pt}
\renewcommand{\arraystretch}{1}
\MakeTableee{|l|cccc|l|}{12cm}{Parameters of the LMC RRL stars modeling from literature}
{
\hline
\douprule
Reference                                            & $b/a$           & $c/a$           & $i\ [{\rm deg}]$      & P.A. $[{\rm deg}]$      & Data \\
\hline
&&&&&\\
\multirow{3}{*}{Pejcha and Stanek (2009)}            & $2.00$          & $3.50$          & $6$            & $113.4$         & OGLE-III RRab \\
                                                     & $1.36$          & $3.53$          & $3$            & $-$             & Removed RRab outside 250 per square degree contour \\
                                                     & $1.99$          & $3.14$          & $9$            & $-$             & Additional color cut \\ 
&&&&&\\
\hline
&&&&&\\
\multirow{2}{*}{Subramaniam and Subramanian (2009)}  & $-$             & $-$             & $31.3\pm3.5$   & $125\pm17$      & OGLE-III RRL stars on-sky projection \\
                                                     & $-$             & $-$             & $20.8\pm3.5$   & $-$             & Included extra-planar features \\ 
&&&&&\\
\hline
&&&&&\\
\multirow{3}{*}{Haschke \etal (2012a)}               & $-$             & $-$             & $32\pm4$       & $114\pm13$      & OGLE-III RRab on-sky projection \\
                                                     & $-$             & $-$             & $-$            & $102\pm21$      & Innermost $3\arcd$ from optical center \\
                                                     & $-$             & $-$             & $-$            & $122\pm32$      & RRL stars $\in(3\arcd,7\arcd)$ from optical center \\ 
&&&&&\\
\hline
&&&&&\\
\multirow{2}{*}{Deb and Singh (2014)}                & $1.67$          & $4.07$          & $24.20$        & $176.01$        & OGLE-III RRab \\
                                                     & $-$             & $-$             & $36.43$        & $149.08$        & OGLE-III RRab plane fitting $|z|=10$ kpc \\ 
&&&&&\\
\hline
&&&&&\\
van der Marel and Kallivayalil (2014)                & $-$             & $-$             & $34.0\pm7.0$   & $139.1\pm4.1$   & Proper motions + old pop. LOS velocity \\
&&&&&\\
\hline
&&&&&\\
This work: $\log(n)=2.15{-} 2.2\ {\rm kpc}^{-3}$ & $1.168\pm0.006$ & $1.950\pm0.022$ & $7.03\pm0.37$  & $19.57\pm2.79$  & \multirow{3}{*}{OGLE-IV RRab} \\
This work: $\log(n)=1.3 {-} 1.4\ {\rm kpc}^{-3}$  & $1.199\pm0.011$ & $1.453\pm0.015$ & $26.22\pm1.34$ & $-30.21\pm3.20$ & \\
This work: $\log(n)=0.0 {-} 0.4\ {\rm kpc}^{-3}$  & $1.250\pm0.018$ & $1.378\pm0.020$ & $36.61\pm3.37$ & $-57.32\pm5.21$ & \\ 
&&&&&\\
\hline
\noalign{\vskip9pt}
\multicolumn{6}{p{12cm}}{For comparison with other tracers see Table~7 in Paper~I.}
}
\end{centering}
\end{landscape}

\noindent
is well correlated with the literature values, not only for the RRL
stars but also for other tracers (see Table~7 in Paper~I). The
position angle is slightly correlated only for larger ellipsoids.

Fig.~4 from Pejcha and Stanek (2009) shows a bar-like structure, that
seems to emerge from the center of the LMC and is elongated along the
line-of-sight (along the $z$ axis). Other studies showed that there is
an evident overdensity in the LMC center (Subramaniam and Subramanian
2009, Haschke \etal 2012a). Fig.~2 from Haschke \etal (2012a) also
seems to show that this overdensity is elongated along the
line-of-sight and forms a bar-like structure (see Fig.~5 in
Haschke \etal 2012a where the RRL stars in the closer bins seem to
form the bar). Subramaniam and Subramanian (2009) state that this RRL
bar-like structure may also aid understanding the LMC bar evolution
suggesting that there must have been a prominent star formation
episode that led to the formation of the LMC disk. Moreover, that
study suggested that the LMC RRL stars were formed in the disk rather
than in the halo.

Our analysis sheds new light on these conclusions based on the central
LMC regions. Because the LMC blend-artifact is very prominent and hard
to remove, and was not easily distinguishable within the OGLE-III
data, it may have been mistakenly treated as the LMC bar. We argue
that the LMC RRL stars distribution does not have a bar, or if there
is one, it is not as prominent as previously thought and a very
careful analysis is needed to extract it from the crowded central
areas of the galaxy.

Subramaniam and Subramanian (2009) obtained the inclination and
position angle of their RRL stars sample very similar to that of the
LMC disk and concluded that most of the LMC RRL stars constitute a
non-spherical structure, while the rest form an inflated
structure. This double-structured RRL stars distribution was later
confirmed by Deb and Singh (2014) based on the metallicity analysis of
the LMC RRL stars. They found that the RRL stars form the disk and the
inner halo. The LMC RRL stars inner halo was also suggested by
Subramanian and Subramaniam (2009). Our analysis of the
three-dimensional distribution of the RRL stars does not support these
findings. Similarly as Pejcha and Stanek (2009) and Haschke \etal
(2012a), we do not see any extra-planar substructures toward
north-east that could be an extension of the disk.  On the other hand,
change in the elongation between the innermost and outermost
ellipsoids may reflect the double nature of the LMC RRL stars
distribution (the disk and the inner halo), but our innermost
ellipsoids are not disk-like (see Fig.~9). Again, the elongation of
the central ellipsoids along the line-of-sight may be affected by
residual blends in our data.

\vspace*{-5pt}
\section{The Small Magellanic Cloud}
\vspace*{-5pt}
\subsection{Three-Dimensional Structure}
\vspace*{-5pt}
In the case of the SMC, RRL stars density in the center is much lower,
so crowding and blending effects are mild, allowing us to study the
galaxy's central regions in detail and compare our results with the
literature. Similarly as its larger neighbor, the SMC also has a
regular, ellipsoidal or nearly spheroidal shape (Kapakos \etal 2011,
Subramanian and Subramaniam 2012, Kapakos and Hatzidimitrou 2012,
Haschke \etal 2012b, Deb \etal 2015). In this section, we concentrate
on the three-dimensional analysis of the SMC using the OGLE-IV
Collection of RRL stars which, in contrast to the OGLE-III Catalog,
covers a very extended area around the SMC (see upper panel in Fig.~5
where the OGLE-IV fields sky coverage and the SMC are presented).
\begin{figure}[bth]
\centerline{\includegraphics[width=12.1cm]{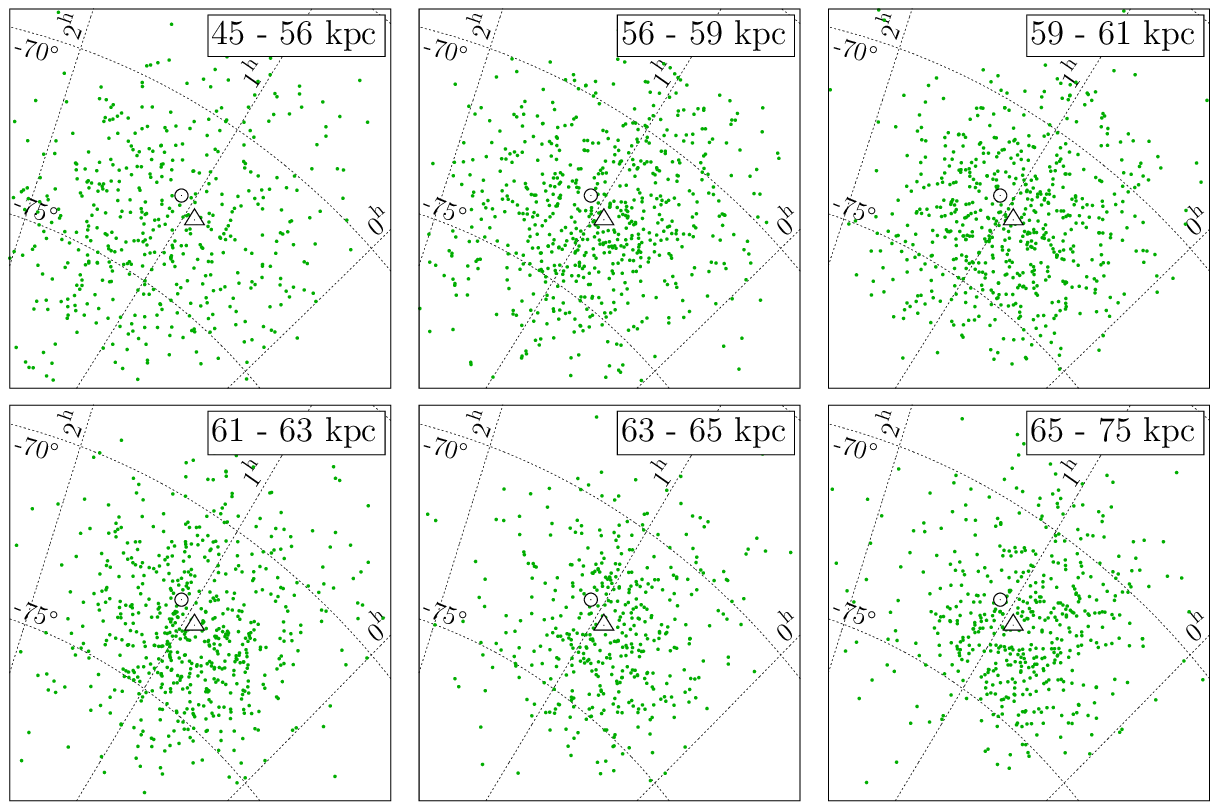}}
\FigCap{Distance tomography of the RRL stars distribution in the SMC in
the Hammer equal-area projection. Note different distance
ranges. White circle marks the SMC dynamical center. White triangle
marks the RRL stars distribution maxima along the RA and Dec axes.}
\end{figure}

Our data show that the SMC has a very regular shape in
three-dimensions (see Fig.~4). Also, the on-sky projection of the SMC
does not present any evident irregularities (see Fig.~5). We decided
to slice-up this galaxy in distance bins in order to see its genuine
structure along the line-of-sight. The distance tomography is shown in
Fig.~11. White circle shows the SMC dynamical center
(Stanimiroviæ \etal 2004) while white triangle shows the SMC RRL stars
distribution center. The latter was estimated in three dimensions
using the maxima of the Right Ascension, Declination and distance RRL
stars distribution which are $\tilde{\alpha}_{\rm
SMC}=0\uph55\upm48\zdot\ups0$, $\tilde{\delta}_{\rm
SMC}=-72\arcd46\arcm48\arcs$, $\tilde{d}_{\rm SMC}=60.45$~kpc. The
median SMC RRL stars distance based on our data is $d_{\rm
SMC,med}=60.58$~kpc. The on-sky distribution center parameters are
significantly shifted with respect to the dynamical SMC center which
are: $\alpha_{{\rm SMC-cen}}=1\uph05\upm$, $\delta_{{\rm
SMC-cen}}=-72\arcd25\arcm12\arcs$ (Stanimiroviæ \etal 2004). The
distribution distance maximum and the median RRL stars distance are
also different from the mean SMC distance obtained from eclipsing
binaries by Graczyk \etal (2014), which is $d_{\rm
SMC}=62.1\pm1.9$~kpc.

\begin{figure}[htb]
\includegraphics[width=13cm]{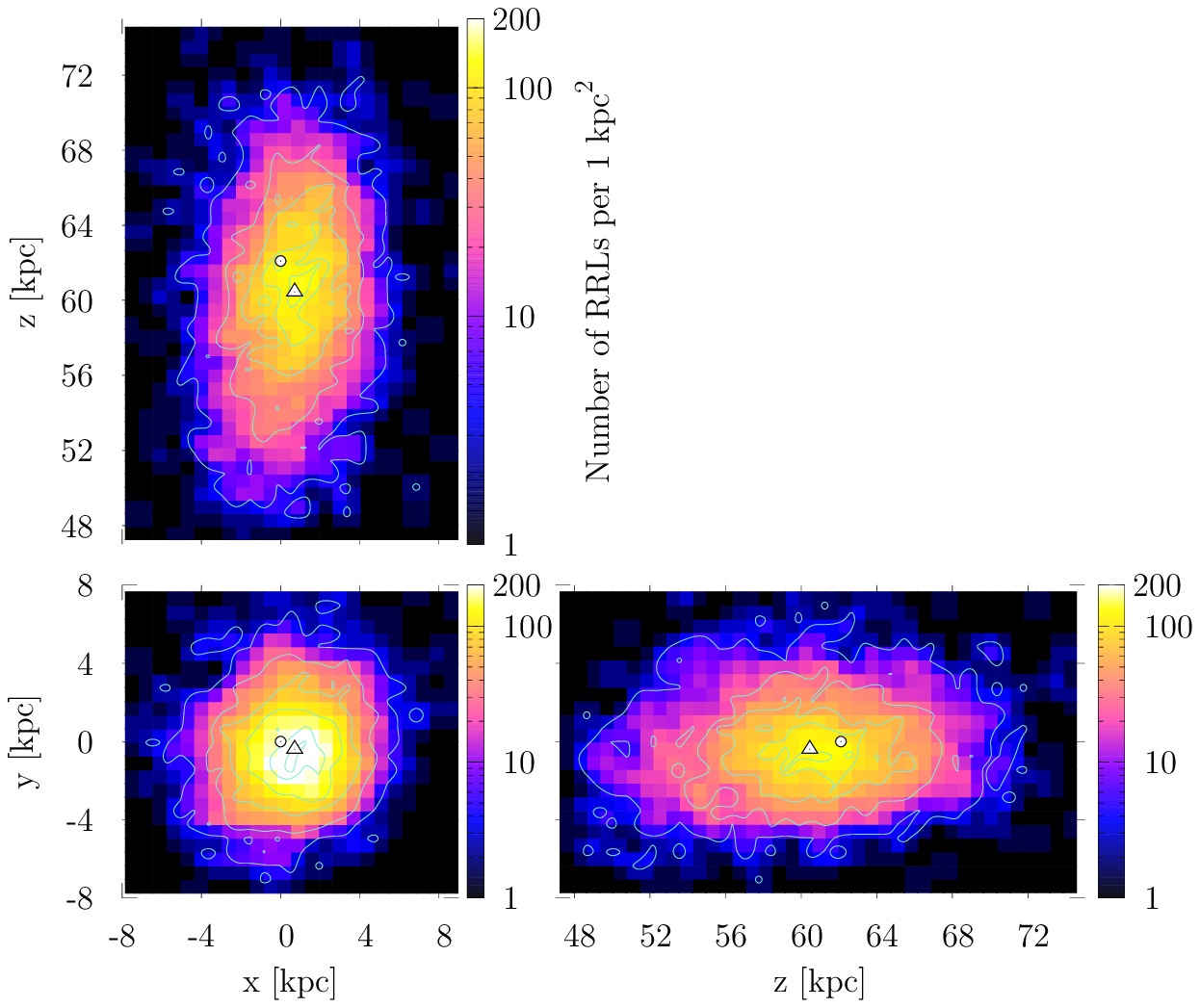}
\FigCap{RRL stars column density maps in the SMC in the Cartesian 
coordinates (the $z$ axis is pointing toward the SMC dynamical
center). Bin size is $0.7$~kpc in $x$, $y$, and $z$ axis. Contour
levels on the $xy$ plane are 5, 30, 70, 120, 200, 260, on the $xz$ and
$yz$ 5, 30, 60, 100, 130 RRL stars per 1~kpc$^2$. White circle and
triangle mark the SMC dynamical and distribution centers,
respectively.}
\end{figure}

The closest RRL stars in the SMC are spread evenly on the sky -- this
is shown in the first panel of Fig.~11. Next three panels presenting
RRL stars around the SMC mean distance do not suggest any asymmetries
or substructures. Last two panels showing the most distant SMC RRL
stars reveal that they are slightly more numerous in the south-western
part of the galaxy than in the north-eastern part.

Fig.~12 shows RRL stars distribution in three dimensions. Bottom left
panel shows the SMC as a regularly, near spheroidally shaped
galaxy. Soszyñski \etal (2010, see their Fig.~7) and Haschke \etal
(2012b, see their Fig.~1) noticed that there are two overdensities in
the SMC center, on-sky projection. A similar feature is visible in the
on-sky projection in the OGLE-IV data (see Fig.~16), but it is not
seen in the three-dimensional Cartesian column density maps (see
Fig.~12). Thus this may be a projection effect. Views ``from the top''
($xz$ plane) and ``from the side'' ($yz$ plane) demonstrate an
elongation of the SMC. This galaxy is stretched almost along the
line-of-sight and its shape is ellipsoidal. No substructures or
evident irregularities can be derived from Fig.~12.

\subsection{Ellipsoid Fitting}
As a result of the analysis from Section~5.1, we decided to model the
SMC RRL stars distribution as a triaxial ellipsoid. The details of the
fitting procedure are given in Section~3.4. We divided the SMC RRL
stars into eleven bins consisting of 126 to 356 stars. The detailed
results of the modeling are presented in Table~5 and in Figs.~13, 14,
and 15.
\begin{figure}[htb]
\includegraphics[width=13cm]{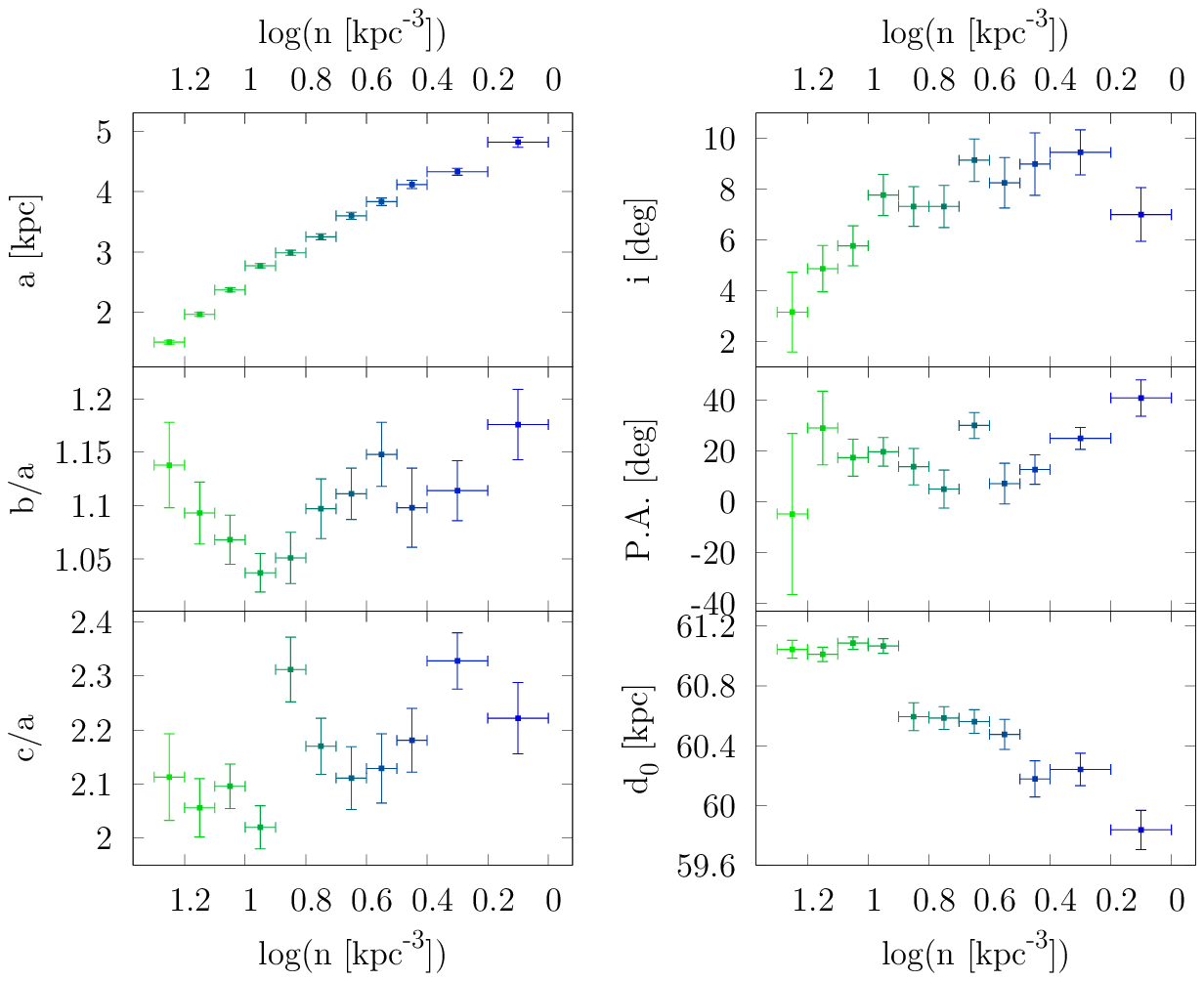}
\FigCap{Parameters of the best-fit triaxial ellipsoids for the SMC RRL
stars. Green points represent the innermost ellipsoids while blue
points -- the outermost.}
\end{figure}

\renewcommand{\arraystretch}{1.3}
\MakeTableSepp{c|ccc}{12cm}{Triaxial ellipsoid best-fit parameters for the SMC}
{
\hline
\noalign{\vskip3pt}
$\log\left(n\left[{\rm kpc}^{-3}\right]\right)$ & $a$ [kpc] & $b/a$ & $c/a$ \\
\noalign{\vskip3pt}
\hline
\noalign{\vskip3pt}
$1.2{-}1.3$ & $1.510\pm0.031$ & $1.138\pm0.040$ & $2.113\pm0.080$ \\ 
$1.1{-}1.2$ & $1.969\pm0.033$ & $1.093\pm0.029$ & $2.056\pm0.054$ \\ 
$1.0{-}1.1$ & $2.375\pm0.034$ & $1.068\pm0.023$ & $2.096\pm0.041$ \\ 
$0.9{-}1.0$ & $2.773\pm0.035$ & $1.037\pm0.018$ & $2.020\pm0.040$ \\ 
$0.8{-}0.9$ & $2.987\pm0.042$ & $1.051\pm0.024$ & $2.312\pm0.060$ \\ 
$0.7{-}0.8$ & $3.253\pm0.049$ & $1.097\pm0.028$ & $2.170\pm0.052$ \\ 
$0.6{-}0.7$ & $3.600\pm0.059$ & $1.111\pm0.024$ & $2.111\pm0.058$ \\ 
$0.5{-}0.6$ & $3.832\pm0.063$ & $1.148\pm0.030$ & $2.129\pm0.064$ \\ 
$0.4{-}0.5$ & $4.117\pm0.069$ & $1.098\pm0.037$ & $2.181\pm0.059$ \\ 
$0.2{-}0.4$ & $4.328\pm0.058$ & $1.114\pm0.028$ & $2.328\pm0.052$ \\ 
$0.0{-}0.2$ & $4.817\pm0.083$ & $1.176\pm0.033$ & $2.222\pm0.066$ \\
\hline
\noalign{\vskip3pt}
$\log\left(n\left[{\rm kpc}^{-3}\right]\right)$ & $i$ [deg] & P.A. [deg]& \\
\noalign{\vskip3pt}
\hline
$1.2{-}1.3$ & $3.16\pm1.57$ &$-4.82\pm31.68$ & \\ 
$1.1{-}1.2$ & $4.87\pm0.91$ &$29.00\pm14.44$ & \\ 
$1.0{-}1.1$ & $5.77\pm0.79$ & $17.33\pm7.29$ & \\ 
$0.9{-}1.0$ & $7.77\pm0.81$ & $19.65\pm5.62$ & \\ 
$0.8{-}0.9$ & $7.32\pm0.78$ & $13.82\pm7.13$ & \\ 
$0.7{-}0.8$ & $7.32\pm0.83$ & ~$5.00\pm7.49$ & \\ 
$0.6{-}0.7$ & $9.14\pm0.84$ & $29.97\pm5.10$ & \\ 
$0.5{-}0.6$ & $8.25\pm0.99$ & ~$7.13\pm8.03$ & \\ 
$0.4{-}0.5$ & $8.99\pm1.23$ & $12.68\pm5.80$ & \\ 
$0.2{-}0.4$ & $9.45\pm0.89$ & $24.91\pm4.28$ & \\ 
$0.0{-}0.2$ & $7.00\pm1.06$ & $40.77\pm7.15$ & \\
\hline
\noalign{\vskip3pt}
${\rm log} \left(n\ \left[{\rm kpc}^{-3}\right]\right)$ & $\alpha_0\ [{\rm deg}]$ & $\delta_0\ [{\rm deg}]$ & $d_0\ [{\rm kpc}]$ \\
\noalign{\vskip3pt}
\hline
$1.2{-}1.3$ & $13.452\pm0.115$ & $-72.987\pm0.023$ & $61.045\pm0.060$ \\ 
$1.1{-}1.2$ & $13.581\pm0.084$ & $-72.993\pm0.026$ & $61.011\pm0.048$ \\ 
$1.0{-}1.1$ & $13.534\pm0.085$ & $-72.985\pm0.023$ & $61.086\pm0.041$ \\ 
$0.9{-}1.0$ & $13.320\pm0.095$ & $-72.958\pm0.026$ & $61.067\pm0.049$ \\ 
$0.8{-}0.9$ & $13.951\pm0.113$ & $-73.000\pm0.036$ & $60.594\pm0.093$ \\ 
$0.7{-}0.8$ & $14.009\pm0.116$ & $-72.985\pm0.035$ & $60.585\pm0.076$ \\ 
$0.6{-}0.7$ & $14.068\pm0.133$ & $-72.894\pm0.037$ & $60.561\pm0.079$ \\ 
$0.5{-}0.6$ & $13.929\pm0.149$ & $-72.874\pm0.047$ & $60.475\pm0.101$ \\ 
$0.4{-}0.5$ & $14.427\pm0.221$ & $-73.048\pm0.065$ & $60.177\pm0.121$ \\ 
$0.2{-}0.4$ & $14.697\pm0.187$ & $-72.876\pm0.050$ & $60.240\pm0.109$ \\ 
$0.0{-}0.2$ & $14.727\pm0.195$ & $-72.877\pm0.070$ & $59.836\pm0.131$ \\
\hline
}

From Table~5 and Fig.~13 we see that for ellipsoids with decreasing
$\log(n)$ (increasing $a$ axis size) both $b/a$ and $c/a$ ratios
neither increase nor decrease and do not change significantly.  This
means that all ellipsoids have virtually the same shape. The median
axis ratio is $1:1.10:2.13$. The inclination angle appears to slightly
decrease from $9\arcd$ to $3\arcd$ in the central regions of the
SMC. Because the inclination is small, the position angle (P.A.) of
the major axis is not well-defined, varying from $-5\arcd$ to
$41\arcd$.

\begin{figure}[htb]
\includegraphics[width=12.5cm]{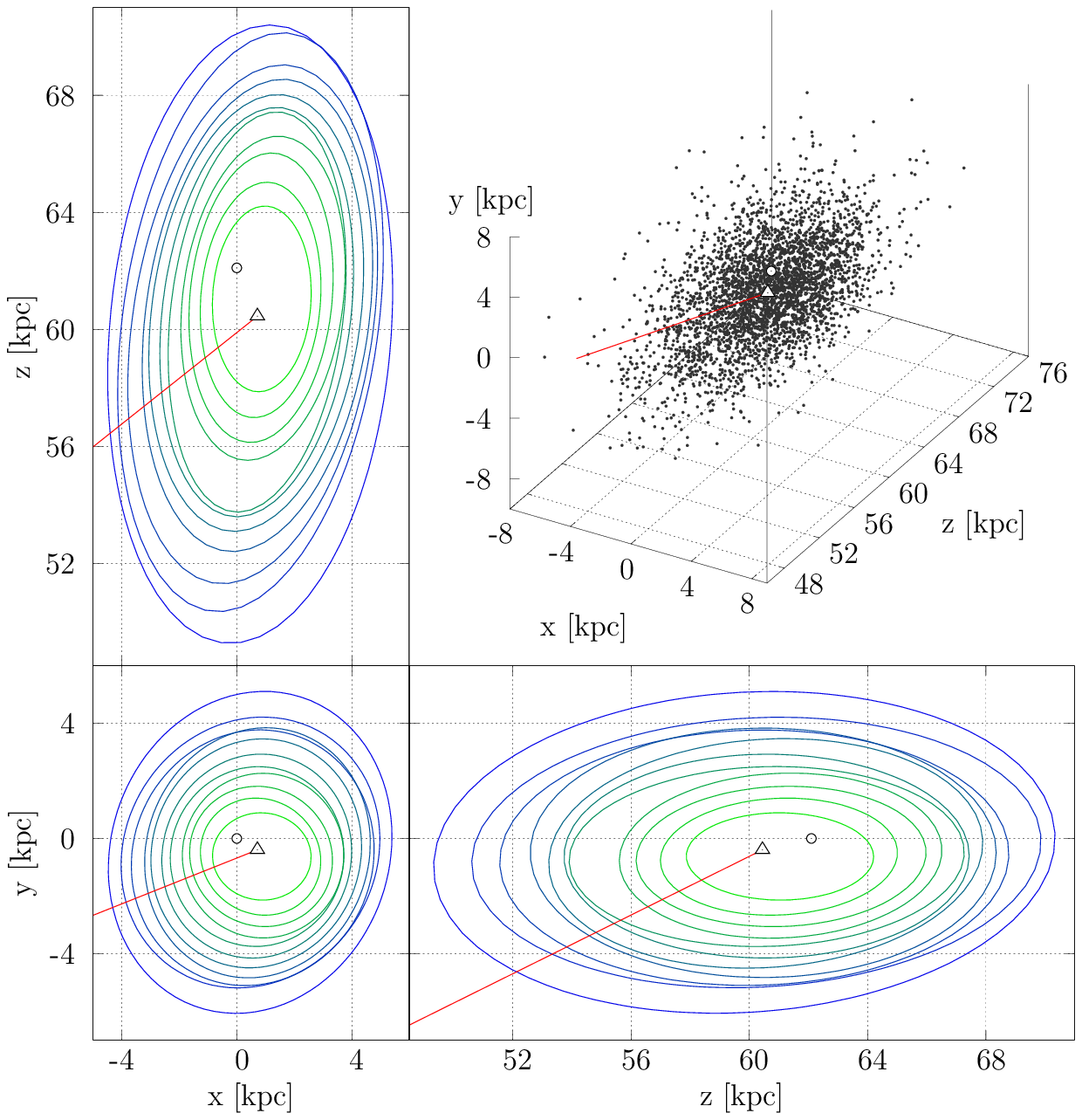}
\FigCap{Best-fit triaxial ellipsoids for the SMC data. Colors are 
compatible with Fig.~13. White circle and triangle mark the SMC
dynamical and distribution centers, respectively. Red line connects
LMC and SMC distribution centers.}
\end{figure}

Fig.~14 shows a three dimensional Cartesian space projections of the
SMC ellipsoids. Both $xy$ and $xz$ planes suggest that the outer parts
of the SMC are more rotated toward the LMC than the inner parts,
although the difference is not very significant and is not visible on
a $yz$ plane. The SMC ellipsoids are elongated almost along the
line-of-sight, as already shown in Fig.~12. Moreover, rotation of
larger ellipsoids on the $xy$ plane toward the LMC may also suggest
that there is an overdensity located near the SMC Wing.

The Cartesian space projections of the ellipsoid centers are shown in
Fig.~15. Green points denote the smallest ellipsoids while blue --
the largest. It is clearly visible that the larger the ellipsoid is
the closer its center is located to the observer (see also Table~5 and
Fig.~13). Moreover, with increasing $a$ axis size the Right Ascension
of the ellipsoid center rises while the Declination does not show
tendency to increase or decrease distinctly. This is reflected in the
Cartesian space projections where centers of larger ellipsoids are
located closer to the LMC. This trend may be caused by the overdensity
in the SMC Wing area or/and the interactions between the Magellanic
Clouds.
\begin{figure}[htb]
\includegraphics[width=13cm]{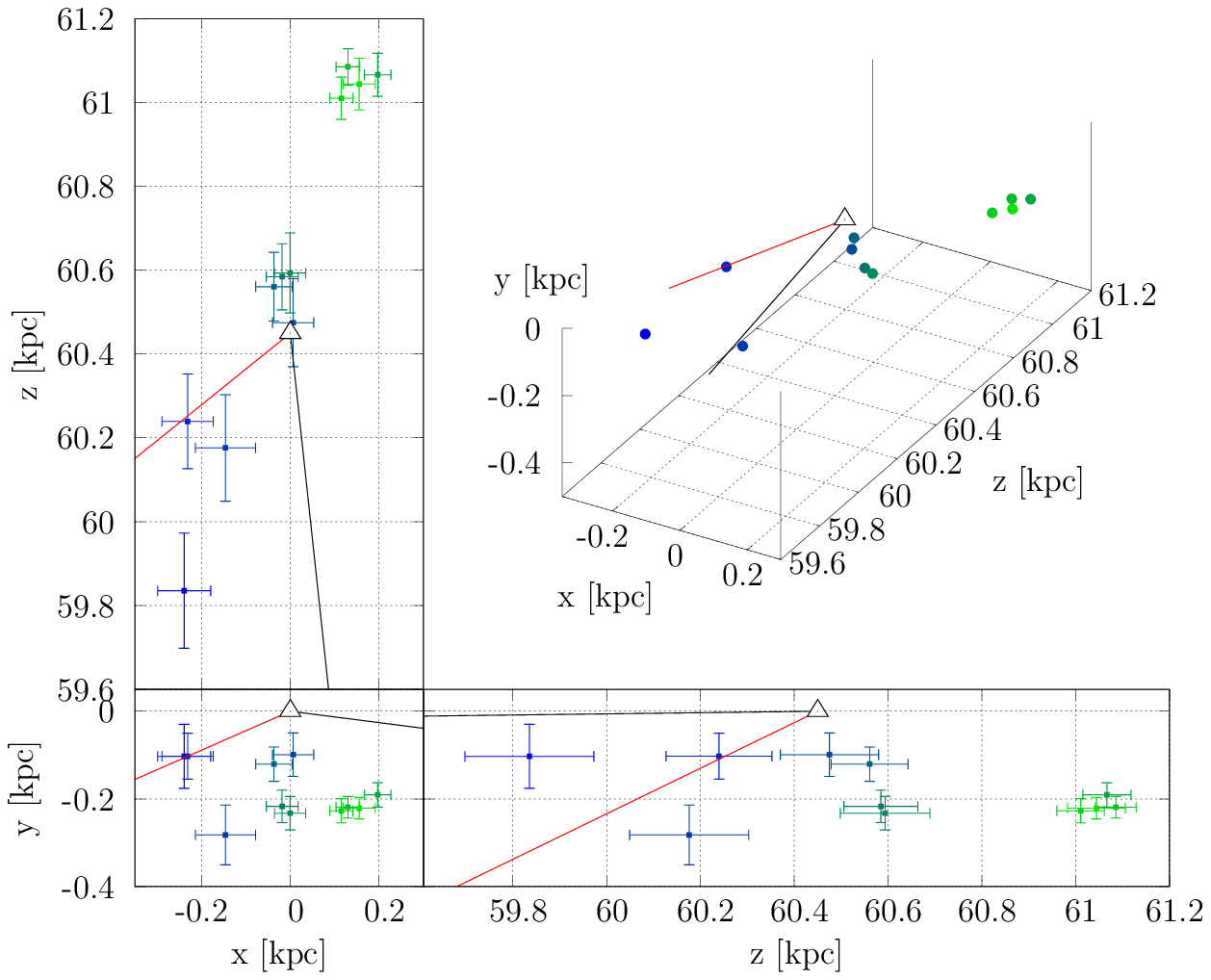}
\FigCap{Best-fit triaxial ellipsoids centers in the Cartesian coordinates
projections for the SMC data. Colors are compatible with Fig.~13 and
14. White triangle marks the RRL stars distribution center. Red line
connects the LMC and SMC distribution centers and black line connects
the SMC center with the Milky Way center (Boehle \etal 2016).}
\end{figure}

\subsection{Comparison with Previous Studies}
A comparison between results obtained in this work and in other
studies is presented in Table~6. Our value of $b/a$ ratio is quite
well compatible with those calculated for the OGLE-III RRL stars
data. The best correlation is for restricted samples (\ie RRL stars
within $r<0\zdot\arcd75$ in Subramanian and Subramaniam 2012 or the
SMC main body in Deb \etal 2015). The differences are caused by the
limited OGLE-III sky coverage. On the other hand, $c/a$ ratio is not
that well correlated. The closest values were also the ones obtained
for restricted samples (\ie RRL stars located within equal extent in
$x$, $y$, and $z$ in Subramanian and Subramaniam 2012 or within
spherical cells in Kapakos and Hatzidimitrou 2012). Other values
suggested very elongated ellipsoids. This is probably again due to the
smaller area observed by OGLE-III.

We also compare tilt parameters in Table~6. The inclination angle
calculated for the OGLE-IV data is compatible with values obtained for
the OGLE-III data. These values fall into the range
$0\arcd{-}7\arcd$. As we have already mentioned, small value

\begin{landscape}
\renewcommand{\arraystretch}{0.9}
\MakeTableee{|l|cccc|l|}{12cm}{Parameters of the SMC RRL stars modeling from literature}
{
\hline
\douprule
Reference                                            & $b/a$           & $c/a$           & $i$ [deg]       & P.A. [deg]       & Data \\
\hline
&&&&&\\
\multirow{9}{*}{Subramanian and Subramaniam (2012)}  & $1.17$          & $1.28$          & $4.2$           & $67.5$           & OGLE-III RRL stars equal extent in $x$, $y$ and $z$: $r<2\zdot\arcd0$ \\
                                                     & $1.24$          & $1.39$          & $3.3$           & $69.5$           & Equal extent in $x$, $y$ and $z$: $r<2\zdot\arcd5$ \\
                                                     & $1.33$          & $1.61$          & $2.6$           & $70.2$           & Equal extent in $x$, $y$ and $z$: $r<3\zdot\arcd0$ \\ 
&&&&&\\
                                                     & $1.07$          & $20.01$         & $0.5$           & $48.84$          & $r<0\zdot\arcd75$ \\
                                                     & $1.30$          & $8.00$          & $0.1$           & $64.87$          & $r<2\zdot\arcd00$ \\
                                                     & $1.33$          & $6.47$          & $0.3$           & $74.40$          & $r<3\zdot\arcd00$ \\ 
&&&&&\\
                                                     & $1.05$          & $19.84$         & $0.4$           & $78.83$          & Excluded 3 NW fields, $r<0\zdot\arcd75$ \\
                                                     & $1.34$          & $8.21$          & $0.1$           & $66.00$          & Excluded 3 NW fields, $r<2\zdot\arcd00$ \\
                                                     & $1.57$          & $7.71$          & $0.4$           & $65.96$          & Excluded 3 NW fields, $r<3\zdot\arcd00$ \\ 
&&&&&\\
\hline
&&&&&\\
Haschke \etal (2012b)                                & $-$             & $-$             & $7\pm15$        & $83\pm21$        & OGLE-III RRab on-sky projection \\ 
&&&&&\\
\hline
&&&&&\\
\multirow{3}{*}{Kapakos and Hatzidimitrou (2012)}    & $1.21$          & $1.57$          & $-$             & $-$              & OGLE-III RRab within spherical cell 2.5 kpc \\
                                                     & $1.18$          & $1.53$          & $-$             & $-$              & Within spherical cell 3 kpc \\
                                                     & $1.23$          & $1.80$          & $-$             & $-$              & Within spherical cell 3.5 kpc \\ 
&&&&&\\
\hline
&&&&&\\
\multirow{2}{*}{Deb \etal (2015)}                    & $1.310\pm0.029$ & $8.269\pm0.934$ & $2.265\pm0.784$ & $74.307\pm0.509$ & OGLE-III RRab \\
                                                     & $1.185\pm0.001$ & $9.411\pm0.860$ & $0.507\pm0.287$ & $55.966\pm0.814$ & The SMC main body \\                                      
&&&&&\\
\hline
&&&&&\\
This work: $\log(n)=1.2{-}1.3~{\rm kpc}^{-3}$  & $1.138\pm0.040$ & $2.113\pm0.080$ & $3.16\pm1.57$   & $-4.82\pm31.68$  & \multirow{3}{*}{OGLE-IV RRab} \\
This work: $\log(n)=0.7{-}0.8~{\rm kpc}^{-3}$  & $1.097\pm0.028$ & $2.170\pm0.052$ & $7.32\pm0.83$   & ~$5.00\pm7.49$   & \\
This work: $\log(n)=0.0{-}0.2~{\rm kpc}^{-3}$  & $1.176\pm0.033$ & $2.222\pm0.066$ & $7.00\pm1.06$   & $40.77\pm7.15$   & \\ 
&&&&&\\
\hline
}
\end{landscape}
\noindent
of $i$ makes P.A. not well defined and we should not rely on a comparison
of this parameter. Even though, the P.A. derived from our sample seems to
be smaller than the ones from the OGLE-III RRL stars.

We do not see any indicators of a bulge or a bar, similarly to
Subramanian and Subramaniam (2012) and Haschke \etal (2012b). Our
equal-density ellipsoids based on the OGLE-IV data that cover a very
extended area around the SMC are all elongated along the line-of-sight
and have almost the same axis ratio. This means that the shape of the
distribution does not change with distance from the center (see
Fig.~14). Thus the elongation along the line-of-sight and so the
higher line-of-sight depth might not indicate the presence of a bulge
as Deb \etal (2015) stated, and as Subramanian and Subramaniam (2009)
deduced from their analysis of the red clump and RRL stars depth
profile.

Many studies revealed that the north-eastern part of the SMC is
located closer to us than the SMC main body (Subramanian and
Subramaniam 2012, Haschke \etal 2012b, Deb \etal 2015). Our data do
not support this as we do not see any irregularities in the SMC
structure that may cause a difference in the mean distance between
some part of this galaxy and the rest (see \ie Fig.~12). This may be
caused by the extended OGLE-IV sky coverage in comparison to the
OGLE-III. On the other hand, we do see some asymmetries of the
equal-density contours (Figs.~12 and~14) that may cause such effect.

\Section{The Magellanic Bridge}
We do see some RRL stars located between the Magellanic Clouds (see
Figs.~4 and 5), although they seem to belong to the halos of the two
galaxies. This is not the first time old stars are observed there
(Bagheri \etal 2013), although we are the first to show a three
dimensional distribution of an old population in the Magellanic
Bridge, represented by RRL stars. Because of the LMC's halo
irregularities and the OGLE-IV limited sky coverage around the
outskirts of this galaxy that we described above it is very difficult
to statistically analyze the area between the Clouds. That is, it is
practically impossible to separate the Bridge RRL stars from the LMC
and SMC halos without having a good model of the LMC outermost halo,
especially that the density of RRL stars in the MBR area is small and
any deviations from the LMC halo density profile would be lost in the
noise. We can only state that these two halos are overlapping.

\begin{figure}[htb]
\includegraphics[width=6.7cm]{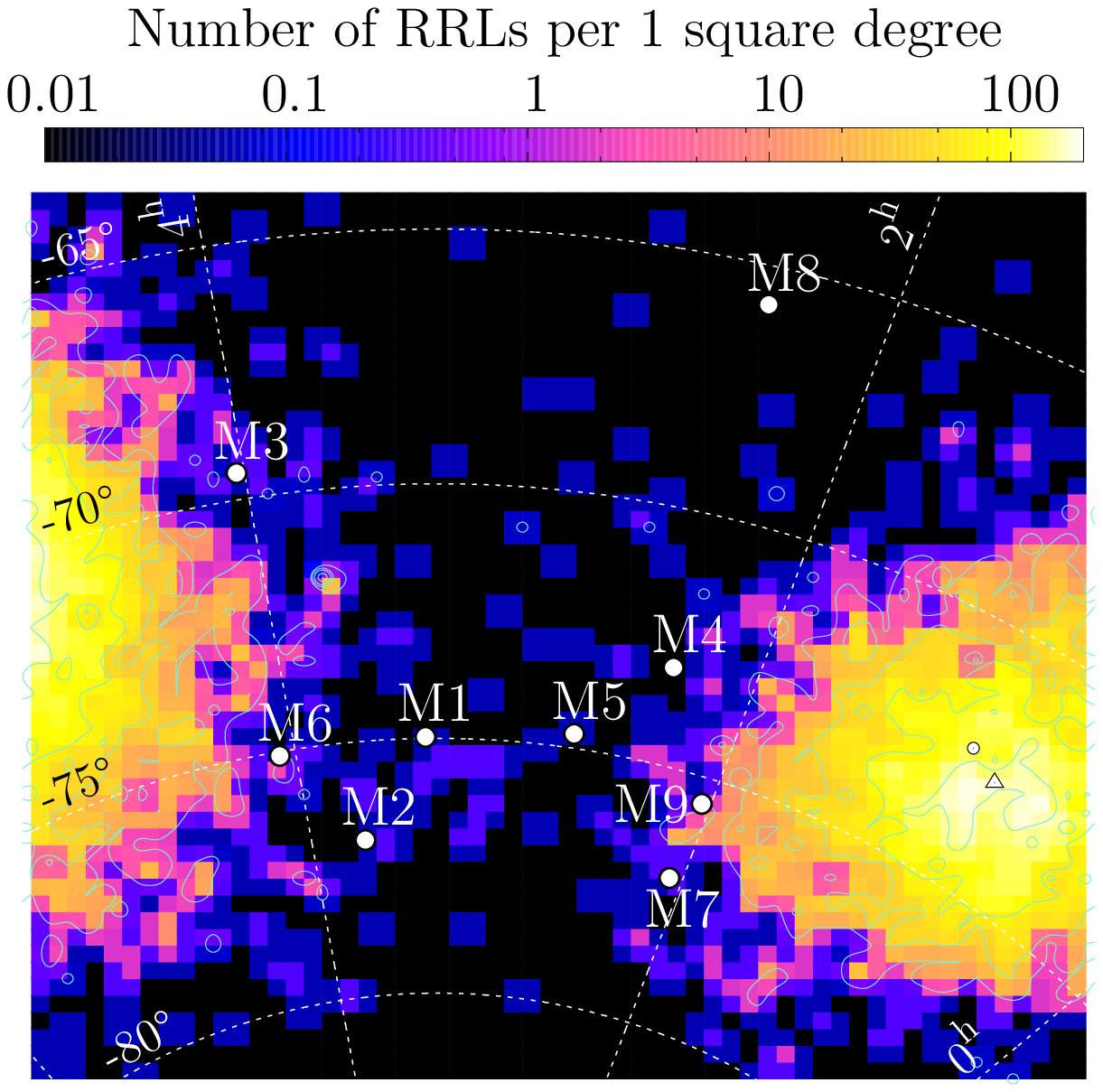}\hfill
\includegraphics[width=6.4cm]{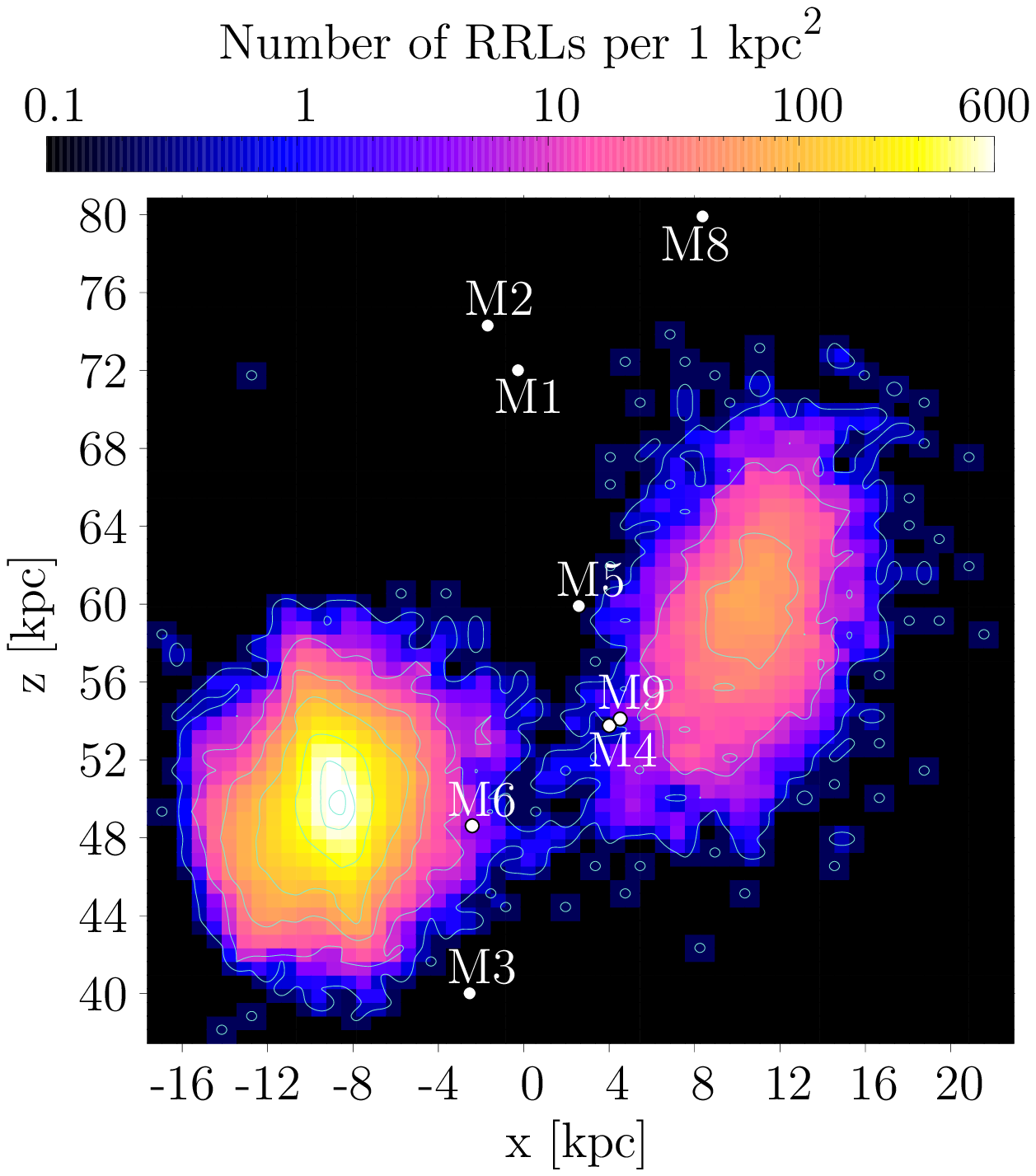}\\
\FigCap{{\it Left panel:} The on-sky projection of the binned RRL stars 
distribution in the Magellanic Bridge area (using Hammer equal-area
projection). The RRL stars column density is color-coded. Additionally,
the Classical Cepheids from Paper~I are marked with white dots. The MBR CCs
are represented with larger dots and labeled M1--M9 as in Paper~I. {\it
Right panel:} The $xz$ plane of the Cartesian projection of RRL stars in
the Magellanic System (view ``from the top''). Bin size is 0.7~kpc in
$x$, $y$, and $z$ axis. Light green lines represent density contours, which
levels are: 1, 10, 40, 100, 300, 600, 700 RRL stars per 1~kpc$^2$.}
\end{figure}

A column density map of the Magellanic Bridge (MBR) is shown in the left
panel of Fig.~16 as an on-sky projection. The RRL stars column density is
color-coded.  The overdensity near the SMC Wing is visible on the right, at
$\alpha\approx2\uph$, $\delta\approx-75\arcd$. There may seem to be an
overdensity connecting the Clouds along ${\rm Dec}\approx-75\arcd$ although
as we have mentioned it is very difficult to analyze this area
statistically and spectroscopic observations will be needed to tell the
true origin of these RRL stars.

Another view of the MBR area is presented in the right panel of Fig.~16. A
column density map of the $xz$ Cartesian space projection shows a ``view
from the top'' of the entire Magellanic System. Additionally, density
contours are plotted with light green lines. Extended SMC halo is fully
pictured while the LMC outskirts reveal limited OGLE-IV sky coverage in the
eastern parts of this galaxy. Without seeing the entire LMC outskirts we
are unable to say if the stars that we see between the Clouds constitute
the genuine MBR. Even though, we can definitely say that the LMC and SMC
halos are overlapping.

\Section{Comparison with Distribution of the Classical Cepheids}
In this section we compare the discussed distribution of the RRL stars with
the distribution of the Classical Cepheids (CCs) that we analyzed in
Paper~I. The RRL stars represent an old stellar population while the CCs
are young stars. Both types of objects in the entire Magellanic System are
shown in Figs.~17 and 18. The former presents data in an on-sky equal-area
Hammer projection, the latter in the three-dimensional Cartesian space
projections.

\begin{figure}[htb]
\includegraphics{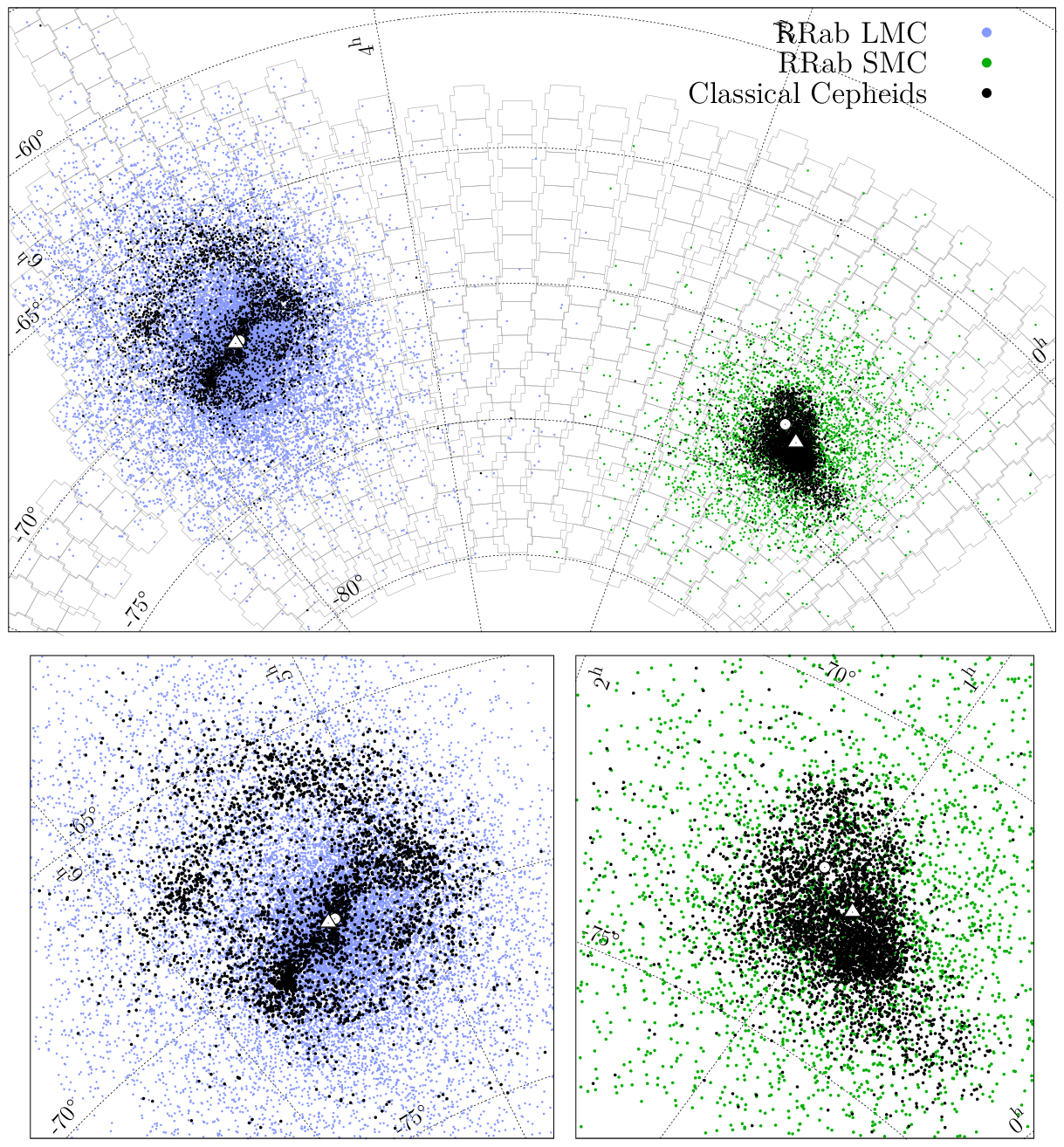}
\FigCap{The equal-area Hammer projection of the RRL stars in the 
Magellanic System -- similar to Fig.~5 but Classical Cepheids from Paper~I
are overplotted with black dots. Blue dots mark the LMC RRL stars and
green dost -- the SMC RRL stars. White circles mark galaxies' dynamical
centers (Stanimiroviæ \etal 2004, van der Marel and Kallivayalil
2014). White triangles mark RRL stars distribution centers.}
\end{figure}

\begin{figure}[htb]
\centerline{\includegraphics[width=12.9cm]{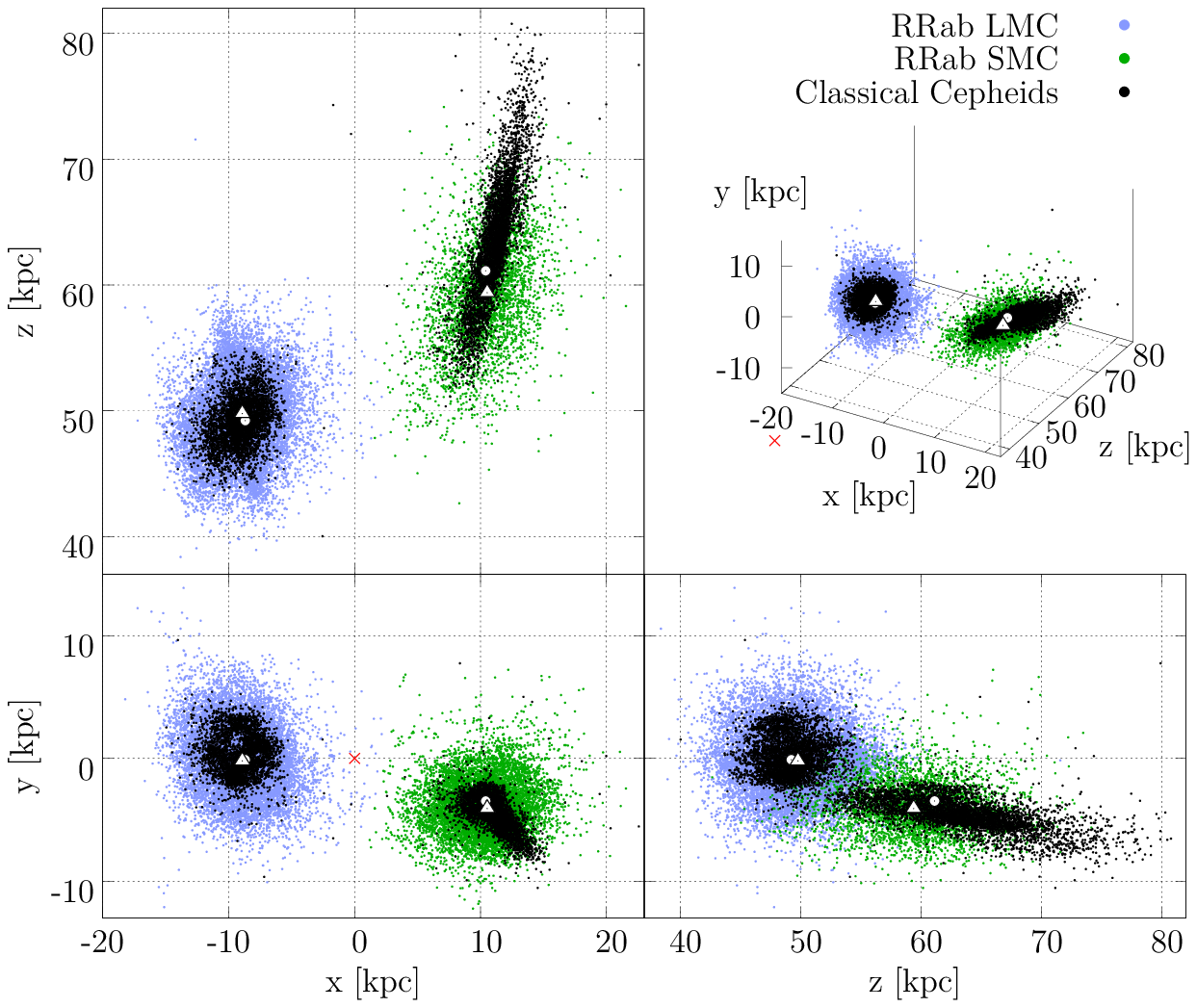}}
\vskip7pt
\FigCap{The RRL stars in the Magellanic System in the Cartesian 
coordinates. The LMC stars are marked with blue dots, while the SMC stars
-- with green dots. Additionally, the Classical Cepheids from Paper~I are
overplotted with black dots. The white circle denotes the LMC
(Pietrzyñski \etal 2013, van der Marel and Kallivayalil 2014) and SMC
(Stanimiroviæ \etal 2004, Graczyk \etal 2014) dynamical centers. White
triangles mark RRL stars distribution centers.}
\end{figure}

\subsection{The Large Magellanic Cloud}
The most obvious difference between the CCs and RRL stars distributions in
the LMC is their spread in the on-sky projection (see Fig.~17). The CCs are
less spread than the RRL stars and are concentrated toward the galaxy
center. The RRL stars are present in every OGLE-IV field and seem to be
located even farther. There are more CCs than the RRL stars in the
northern parts of the inner LMC, because of the well populated northern arm
of this galaxy. The on-sky projection in Fig.~17 also shows that the CCs
are located mainly in the LMC substructures: the bar and northern arm. The
RRL stars are distributed definitely more smoothly and regularly and we do
not see any evident substructures. The CCs distribution in the LMC can be
modeled with a plane (see Paper~I), while the RRL stars distribution is
modeled as a triaxial ellipsoid that is far from being flat and so the LMC
RRL stars may not be described as a plane.

The three-dimensional Cartesian space projections in Fig.~18 also show
differences between the CCs and RRL stars distributions. The median
distance of the LMC RRL stars is $d_{\rm RRL,med}=50.64$~kpc, while for the
Cepheids it was $d_{\rm CC,med}=49.93$~kpc (see Table~4 in Paper~I). These
values are in good agreement within distance mean uncertainties and
distance standard deviations, and a similar conclusion was reached by
Haschke \etal (2012a). The $xy$ plane represents a similar view to the
on-sky projection from Fig.~18 that we have described above. View ``from
the top'' ($xz$ plane) again shows that the RRL stars distribution could
not be described properly as a disk. Moreover, the CCs in the LMC were not
as affected by crowding and blending effects (see \ie Fig.~5 in Paper~I) as
the RRL stars. This is probably due to the fact that the RRL stars are
fainter and have higher column density in the LMC center than the CCs. The
$yz$ plane only shows that the RRL stars are more spread than the CCs.

\subsection{The Small Magellanic Cloud}
Similarly as in the LMC, the RRL stars and CCs in the SMC are distributed
differently. Again, older stars are more spread and form a regular
structure in the on-sky projection, while younger stars are more clumped
and concentrated near the galaxy center (see Fig.~17). The CCs seem to be
more numerous in the south-western part of the SMC.

The Cartesian coordinates projections in Fig.~18 show great differences
between the RRL stars and CCs distributions in the SMC. The median distance
of the RRL stars is $d_{\rm RRL,med}=60.58$~kpc and for the CCs it was
$d_{\rm CC,med}=64.62$~kpc (see Table~8 in Paper~I). This time the
difference is larger than for the LMC and these values are not correlated
within median distance uncertainties. Even though, they are within distance
standard deviations. The difference may also be an effect of different
methods of distance calculations for the CCs and RRL stars. The former were
calculated relative to the LMC distance from Pietrzyñski \etal (2013),
assuming the same zeropoint of the P-L relation in both the LMC and SMC,
while the latter were obtained independently of any other distance
estimations. However, other studies show that the mean distance calculated
for the RRLs is smaller than that for the CCs (Haschke \etal 2012b, de
Grijs and Bono 2015) and this is in good agreement with our results.

The $xy$ plane confirms that the RRL stars are more spread and constitute a
very regular shape, while the CCs form a structure that is very
elongated. The $xz$ and $yz$ projections demonstrate the SMC CCs shape that
is stretched along the line-of-sight. In this direction the RRL stars do
not reach that far and are less elongated than CCs, which is reflected in
median distance differences.

\subsection{The Magellanic Bridge}
The RRL stars on-sky column density map of the Magellanic Bridge area
showing also CCs locations is presented in the left panel of Fig.~16. The
Bridge Cepheids are marked with large white dots and labeled M1--M9 (as in
Paper~I). Interestingly, their positions seem to be correlated with
slightly higher RRL stars densities, especially those located along
Declination $\approx-75\arcd$.

A very different picture is presented in the Cartesian coordinates $x$ and
$z$ projection of the same area that is shown in the right panel of
Fig.~16. The Bridge Cepheids are very spread along the $z$ axis (along the
line-of-sight). Only three of them fall into higher RRL stars density
contour at the level of 1~RRL star per~kpc$^2$ (M4, M6, and M9) and two
other are quite close (M3 and M5). The highest number of RRL stars per
1~kpc$^2$ in the Bridge area is reached strictly between the Clouds and we
would expect to find the genuine MBR RRL stars right there. However, even
if we account for the errors in distance estimations, the locations of MBR
CCs and RRL stars situated between the Clouds are not correlated.

\Section{Conclusions}
In this work, we present the analysis based on a sample 19~401 RRab
selected from the newest release of the OGLE Collection of RRL stars in the
Magellanic System (Soszyñski \etal 2016a) based on the OGLE-IV data
(Udalski \etal 2015).

The LMC has a regular shape in three dimensions and no prominent
substructures are distinguishable. Even though, the LMC halo is
slightly asymmetrical with larger number of RRL stars in its
north-eastern part, which is also located closer to us than the entire
LMC. We argue that the putative LMC bar in RRL stars is in fact an
effect of strong blending and crowding effects in the LMC center, and
it was not possible to distinguish before the OGLE-IV extensive data
were available. Triaxial ellipsoids were fitted to surfaces of
constant number density, excluding the densest central region. Smaller
ellipsoids have higher axis ratio and are elongated along the
line-of-sight, which is probably not physical due to the residual
blends. Larger ellipsoids are slightly more rotated toward the SMC
although not entirely. The inclination and position angle change
substantially with the $a$ axis size. The ellipsoid centers move away
from the SMC and from the observer and Milky Way center with
increasing $a$ axis size.

The SMC is mostly free from the blending and crowding effects, due to a
significantly smaller number of RRL stars in this galaxy. The SMC has a
very regular shape in three-dimensions and we do not see any substructures
or asymmetries. We only see a slightly higher column density near the SMC
Wing. The distribution center is very different from the dynamical center,
which was not the case for the LMC. All ellipsoids fitted to surfaces of
constant number density have virtually the same shape (\ie axis
ratios). The inclination angle is very small thus the position angle is not
well defined. In contrary to the LMC, SMC ellipsoids centers move toward
the LMC, the observer and the Milky Way center with increasing $a$ axis
size.

We show, for the first time, a three dimensional distributions of the RRL
stars in the extended area between the Magellanic Clouds -- the Magellanic
Bridge. Unfortunately, we are unable to separate two Clouds' halos from
each other and thus we cannot differentiate the genuine Bridge RRL stars
from those belonging to the LMC or SMC. This is mostly because of the
limited OGLE-IV sky coverage on the eastern side of the LMC. With the LMC
halo being asymmetrical and not fully covered it is very difficult to
analyze the Bridge area statistically, especially that the RRL stars
numbers in the Bridge are small and most probably any deviations from the
LMC/SMC halo profile would be lost in the noise. We can only state that the
Clouds' halos are overlapping.

A comparison with the results from Paper~I clearly shows that the Classical
Cepheids and the RRL stars are distributed differently in both Magellanic
Clouds. The younger stars are clumped and constitute substructures while
the older are more spread and distributed regularly. For the LMC we have
obtained a very similar median distance for the CCs and RRL stars, in
contrary to the SMC, where the difference is $\approx4$~kpc. The CCs
distribution is definitely showing signs of Clouds' interaction, while it
is not easy to find such evidence in the RRL stars distribution. In the
Magellanic Bridge area on-sky projection, CCs seem to be located near the
highest column density of RRL stars between the Clouds. On the other hand,
Cartesian $xz$ projection shows that the three-dimensional correlation is
very small and while the RRL stars are located mainly between the Clouds,
the CCs tend to spread far beyond.

\Acknow{A.M.J.-D. is supported by the Polish Ministry of Science and Higher
Education under ``Diamond Grant'' No. 0148/DIA/2014/43. D.M.S. is supported
by the Polish National Science Center under the grant
2013/11/D/ST9/ 03445 and the Polish Ministry of Science and Higher
Education under the grant ``Iuventus Plus'' No. 0420/IP3/2015/73. The OGLE
project has received funding from the National Science Center, Poland,
grant MAESTRO 2014/14/A/ST9/00121 to AU.}

\end{document}